\documentclass[aps,prc,reprint,showpacs,groupedaddress,onecolumn]{revtex4-1}

\usepackage{graphicx} % Include figure files 
\usepackage{dcolumn}  % Align table columns on decimal point  
\usepackage{bm}       % bold math 
\usepackage{amsmath}

\begin{document}
\newcommand{\vv}[1]{{$\bf {#1}$}}
\newcommand{\ul}[1]{\underline{#1}}
\newcommand{\vvm}[1]{{\bf {#1}}}
\def\bsigma{\mbox{\boldmath$\sigma$}}

\title{Open Shell Effects in a Microscopic Optical Potential for 
Elastic Scattering of $^{6(8)}$He}

\author{A. Orazbayev}
\email {ao379408@ohio.edu}

\author{Ch. Elster}
\email{elster@ohio.edu}

\affiliation{Institute of Nuclear and Particle Physics,  and
Department of Physics and Astronomy,  Ohio University, Athens, OH 45701
}

\author{S.P. Weppner}
\email{weppnesp@eckerd.edu}

\affiliation{ Natural Sciences, Eckerd College, St. Petersburg, FL 33711}

\date{\today}

\begin{abstract}
Elastic scattering observables (differential cross section and analyzing power) are
calculated for the reaction $^6$He(p,p)$^6$He at projectile energies starting at
71~MeV/nucleon. The optical potential needed to describe the reaction 
is based on a microscopic Watson first-order folding potential, which explicitly
takes into account that the two neutrons outside the $^4$He-core occupy an
open p-shell. The folding of the single-particle harmonic oscillator
density matrix with the
nucleon-nucleon t-matrix leads for this case to new terms not present in
traditional folding optical potentials for closed shell nuclei.
The effect of those new terms on the elastic scattering observables is
investigated. Furthermore, the influence of an exponential tail of the p-shell
wave functions on the scattering observables is studied, as well as the
sensitivity of the observables to variations of matter and charge radius. Finally
elastic scattering observables for the reaction $^8$He(p,p)$^8$He are presented
at selected projectile energies.

\end{abstract}


\maketitle

\section{Introduction}

The  exotic helium isotopes have been extensively studied, both experimentally and
theoretically. The charge radii of  $^6$He and  $^8$He are experimentally very well
known~\cite{Wang:2004ze,Mueller:2008bj,Brodeur:2012zz}.
The nucleus $^6$He is of particular interest since it constitutes the lightest
two-neutron halo nucleus with a $^4$He core.   
Investigating 
its structure  already inspired a large body of
work including effective few-body
models~\cite{Lehman:1983zz,Zhukov:1993aw,Fedorov:2003jx}, multi-cluster
methods~\cite{Csoto:1993fg,Varga:1994fu,Wurzer:1996rr} 
Green's Function Monte Carlo (GFMC) methods~\cite{Pudliner:1997ck}, and
no-core shell model calculations~\cite{Navratil:2004dp,Navratil:2009ut,Bacca:2012up}, 
so that ground state properties of $^6$He appear to be quite well understood. 
Similarly, the ground state properties of $^8$He have been explored with different
theoretical methods~\cite{Bacca:2009yk,Itagaki:2008zz}.

Recently, elastic scattering of $^6$He~\cite{Uesaka:2010mm,Hatano:2005} as well
as $^8$He~\cite{Sakaguchi:2013uut}  off a polarized proton target has been measured
for the first time at a laboratory kinetic energy of 71~MeV/nucleon.
The experiments find that for $^6$He the analyzing power $A_y$ becomes negative around
50$^o$, whereas for $^8$He it stays positive.
Specifically the behavior of $A_y$ for $^6$He  not predicted by simple folding 
models for the optical
potentials~\cite{Weppner:2000fi,Gupta:2000bu}, though the calculations
reproduce the differential cross section at this energy reasonably well.

This apparent ``$A_y$ problem" conveys the inadequacy of
using the same
methods which describe p-A scattering from stable nuclei
for reactions involving halo nuclei. The obvious difference is the nuclear
structure. Traditionally, microscopic folding models are developed for closed shell
nuclei, like $^{16}$O,  $^{40}$Ca, or  $^{208}$Pb. Though $^6$He and $^8$He are both
spin-0 nuclei, their outer p-shell is not fully occupied. In the case of $^6$He 
two neutrons occupy the p-shell. This structure suggests describing $^6$He with  
three-body cluster models, as pioneered in 
Refs.~\cite{Crespo:2006vg,Crespo:2007zz} for higher energies.
For describing the differential cross section and the analyzing power at
71~MeV/nucleon, Refs.~\cite{Uesaka:2010mm,Weppner:2011px} 
use  ``cluster-folding" calculations
with still only limited success at understanding the $A_y$ problem.

The focus of this work is to extend traditional microscopic folding models 
to take the valence neutrons in $^{6(8)}$He explicitly into account.
In order to facilitate this calculation, we assume a simple harmonic oscillator model
ansatz for $^{6(8)}$He. In Section~II we derive the formulation for a microscopic
optical potential which takes into account the partially occupied p-shell of
$^{6}$He, and show the resulting effect on the differential cross section and the
analyzing power at different energies.
Since we use a model based on oscillator wave functions, we investigate in Section III,
if this specific functional form of the wave functions has an effect on the scattering
observables at energies of  71~MeV/nucleon and higher. Specifically we study, if there is
a difference at these energies between wave functions that fall off exponentially in
coordinate space or harmonic oscillator wave functions.  In Section IV  we study the
sensitivity of the scattering observables to the charge and matter radii of $^{6}$He. 
In Section V we study the open shell effects in the optical potential on the scattering
observables for $^{8}$He. We conclude in Section VI.

%%%%%%%%%%%%%%%%%%%%%%%%%%%%%%%%%%%%%%%%%%%%%%%%%%%%%%%%%%%%%%%%%

\section{Open Shell Effects in the Optical Potential for $^{6}$He}

\begin{figure}
\begin{center}
\includegraphics[scale=.55]{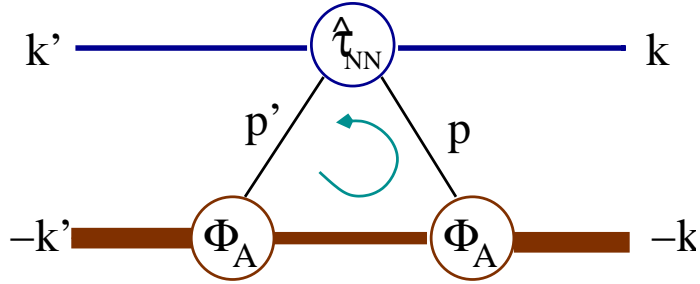}
\caption{Diagram for the  folding optical potential matrix 
element in the single-scattering approximation.
\label{fig1}
}
\end{center}
\end{figure}

Let $H=H_0+V$ be the Hamiltonian for the nucleon-nucleus system in which the
interaction $V=\sum_{i=1}^A v_{0i}$ consists of all two-nucleon interactions $v_{0i}$
between the projectile (``$0$'') and a target nucleon (``$i$''). The free Hamiltonian
is given by $H_0=h_0 +H_A$, where $h_0$ describes the kinetic energy of the
projectile, while the target Hamiltonian $H_A$ satisfies $H_A |\Phi_A\rangle =
E_A|\Phi_A\rangle$, with $|\Phi_A\rangle$ being the ground state of the target.
Focusing on elastic scattering, the transition operator  is
given by 
\begin{equation}
PTP \equiv T_{el} =PUP + PUPG_0(E)PT_{el},
\label{eq:2.1}
\end{equation}
where $P=\frac{|\Phi_A \rangle \langle \Phi_A|}{\langle \Phi_A| \Phi_A \rangle }$ is
the projection operator onto the ground state $|\Phi_A\rangle$ with $P+Q={\bf 1}$,
where $Q$ projects onto the orthogonal space, and  $G_0(E) = (E-h_0 -H_A
+i\varepsilon)^{-1}$ is the propagator, which here will be treated in impulse
approximation. The Watson first-order optical potential operator for scattering 
of protons is given by~\cite{Chinn:1993zza} and Appendix A of Ref.~\cite{Weppner:2011px}
\begin{equation}
U_p= \sum_{i=1}^Z \tau_{0i}^{pp} + \sum_{i=1}^N \tau_{0i}^{np}
\equiv U_p^Z+U_p^N, \label{eq:2.2}
\end{equation}
where the two-body transition operators $\tau^{pp(np)}_{0i}$ are related to the
proton-proton ($pp$) and neutron-proton ($np$) t-matrices $\hat\tau^{pp(np)}_{0i}$
 via~\cite{Chinn:1993zza}
\begin{equation}
\tau^{pp(np)}_{0i} = {\hat\tau}^{pp(np)}_{0i} - {\hat\tau^{pp(np)}_{0i}} G_0(E) P
\tau^{pp(np)}_{0i}.
\label{eq:2.3}
\end{equation}
As function of the external momenta $\vvm k$ and $\vvm k'$ the first-order optical
potential is given by
\begin{equation}
\langle \vvm k' | \langle \phi_A|PUP|\phi_A\rangle
|\vvm k\rangle\equiv U_{el}(\vvm k',\vvm k)=\sum_{i=N,P}\left
\langle \vvm k' | \langle \phi_A|\hat\tau_{0i}({\cal{E}})|\phi_A\rangle
|\vvm k\right\rangle , 
\label{eq:2.4}
\end{equation}
where ${\cal{E}}$ is the energy of the system. In this work the common approximation
of fixing ${\cal{E}}$ at half the laboratory energy will be used. The summation over
$i$ indicates that one has to sum over $N$ neutrons and $Z$ protons.  The structure
of Eq.~(\ref{eq:2.4}) is schematically indicated in Fig.~\ref{fig1}, where $\vvm p$ and
$\vvm p'$ are the internal variables of the struck target nucleon, which enter into
the two-body t-matrices as well as the single-particle densities.

Let us first consider the nucleon-nucleon (NN) t-matrix.
On the energy shell, the NN scattering-amplitude matrix $\overline{M}(\vvm p'_{NN},\vvm
p_{NN})$
is related to the on-shell transition matrix element as $\overline{M}(\vvm p'_{NN},\vvm
p_{NN})= -4 \pi^2
\mu_{NN} \langle \vvm p'_{NN} |\hat{\tau}_{0i}|\vvm p_{NN}\rangle$, where $\mu_{NN}$ is the
reduced mass of the two-nucleon system. The off-shell Wolfenstein~\cite{wolfenstein-ashkin}
 parameterization of $\overline{M}(\vvm p'_{NN},\vvm p_{NN})$ is given by 
\begin{eqnarray}
\overline{M}&=& A I + i C (\bsigma^{(0)} \otimes  I  + I \otimes \bsigma^{(i)})\cdot
\hat{\mathbf{n}}_{NN} 
+ M (\bsigma^{(0)} \cdot \hat{\mathbf{n}}_{NN}) \otimes (\bsigma^{(i)} \cdot
\hat{\mathbf{n}}_{NN})  \cr
& & + (G+H)(\bsigma^{(0)} \cdot \hat{\mathbf{K}}_{NN}) \otimes (\bsigma^{(i)} \cdot
\hat{\mathbf{K}}_{NN}) +
(G-H)(\bsigma^{(0)} \cdot \hat{\mathbf{q}}_{NN}) \otimes (\bsigma^{(i)} \cdot
\hat{\mathbf{q}}_{NN})  \cr
& & + D \left((\bsigma^{(0)} \cdot \hat{\mathbf{q}}_{NN}) \otimes
(\bsigma^{(i)} \cdot \hat{\mathbf{K}}_{NN}) + (\sigma^{(0)} \cdot \hat{\mathbf{K}}_{NN}) \otimes
(\bsigma^{(i)} \cdot \hat{\mathbf{q}}_{NN}) \right)
\label{eq:2.5}
\end{eqnarray} 
The spin-momentum operators of Eq.~(\ref{eq:2.5}) are invariant with respect to
rotations, and spin exchange. They are time reversal invariant with the exception of the
last operator, which changes sign and thus is paired with a coefficient function $D$,
that is odd in $|\vvm p'_{NN}|^2 - |\vvm p_{NN}|^2$, and thus vanishes
on-shell. The Wolfenstein amplitudes are functions of the vector variables $\vvm p'_{NN}$
and $\vvm p_{NN}$ and can be either calculated directly as such~\cite{Veerasamy:2012sp}
or obtained from partial wave sums. The momentum vectors are defined as 
$\vvm q_{NN} = \vvm p'_{NN} - \vvm p_{NN}$, $\vvm K_{NN} = \vvm p'_{NN} + \vvm p_{NN}$,
and $\vvm n _{NN} = \vvm p' _{NN} \times \vvm p _{NN}$, and given in the two-nucleon intrinsic frame.  

For the calculation of the optical potential of Eq.~(\ref{eq:2.4}) the expectation
values of these spin-momentum operators need to be calculated in the plane-wave basis
for the projectile characterized by $\bsigma^{(0)}$ and in a nuclear basis for the
struck nucleon characterized by $\bsigma^{(i)}$.

\subsection{Model for the Single Particle Density of $^{6}$He}
\label{subdenshe6}

Since our goal is to explore the folding optical potential for a nucleus with an open-shell
structure, 
we first need to consider the explicit angular momentum and spin structure of the
single particle density that enters the folding optical potential. Without loss of
generality we assume nucleon ``$1$'' is the struck target nucleon, so that 
\begin{eqnarray}
\rho_{I,M_I;I,M_I} (1,1') &=& 
  \int \prod_{l=2}^{A-1}d{\bf \zeta_l'}
 \int\prod_{j=2}^{A-1}d{\bf \zeta_j} 
\langle \phi_{I,M_I} |\vvm\zeta_1'\vvm\zeta_2'\vvm\zeta_3'\vvm\zeta_4'...
\vvm\zeta_{A-1}'\rangle\;
\langle \vvm\zeta_1\vvm\zeta_2\vvm\zeta_3\vvm\zeta_4...\vvm\zeta_{A-1}
|\phi_{I,M_I'}\rangle \cr
&\equiv& \langle \phi_{I,M_I} | \psi^\dagger(1) \psi(1') |\phi_{I,M_I}\rangle ,
\label{eq:2.5a}
\end{eqnarray}
where $I$ is the total angular momentum of the ground state, and $M_I$ its
projection. All internal variables integrate out, and one is left with an operator 
$\psi^\dagger(1)$ that creates a nucleon with given quantum numbers
``$1$'', e.g. momentum and spin, which can then be expanded in terms of single particle
wave functions $\phi_{nljm}(1)$ as
\begin{equation}
\psi^\dagger (1) =\sum_{nljm} \phi_{nljm}(1) (a_{nljm})^\dagger .
\end{equation}
Expanding the single particle wave function explicitly into spin, orbital angular momentum, and
radial parts leads to 
\begin{eqnarray}
\rho_{I,M_I;I,M'_I} (1,1') &=& \sum
C_{\lambda\hspace{1pt}m_s\hspace{1pt}m}^{l \hspace{3pt}\frac{1}{2}\hspace{6pt}j} 
C_{\lambda'
\hspace{2pt}m'_s\hspace{1pt}m'}^{l'\hspace{3pt}\frac{1}{2}\hspace{6pt}j'} \;
 Y_l^{\lambda}(1) \chi_{m_s}(1) R_{nlj}(1) \; Y_{l'}^{* \lambda'}(1')
\chi^*_{m'_s}(1') R^*_{n'l'j'}(1') \cr
& & \times \langle \phi_{I,M_I} |(a_{nljm})^\dagger a_{n'l'j'm'} |\phi_{I,M'_I} \rangle.
\label{eq:2.5b}
\end{eqnarray} 
Here the sum is taken over all quantum numbers occurring in the sum. 
This expression exhibits the spin
eigenfunctions of the struck nucleon, but is not yet in a form best suited for
evaluation of matrix elements. Let us define a tensor operator $\tau_{k_s,
q_s}(s=\frac{1}{2})$ for which $k_s$~=0 or 1 with
\begin{eqnarray}
\tau_{00} &=&1 \cr
\tau_{10}&=&2 \sigma_z \cr
\tau_{1\pm 1}&=& \frac{1}{\sqrt{2}} {\mp}(\sigma_x \pm i\sigma_y),
\label{eq:2.5c}
\end{eqnarray} 
where $\sigma_i$ are the usual spin-projections. The matrix elements of this
operator can be written as
\begin{equation}
\langle s m_s | \tau_{k_s,q_s}(s) | s m'_s\rangle = \sqrt{2k_s +1} \;
C_{m'_s\hspace{1pt}q_s\hspace{1pt}m_s}^{s \hspace{5pt} k_s\hspace{4pt}s}
\label{eq:2.5d}
\end{equation}
Inserting Eq.~(\ref{eq:2.5d}) into Eq.~(\ref{eq:2.5b}) and re-coupling the angular
momenta leads to
\begin{eqnarray}
\rho_{I,M_I;I,M_I} (1,1') &=& \sum_{k_l,q_l,k_s,q_s,k,q, \cdots} 
{\mathcal N} \langle \phi_{I,M_I}
|(a_{nljm})^\dagger a_{n'l'j'm'} |\phi_{I,M'_I} \rangle \cr
&& (-1)^{j'-m'}\; C_{m\hspace{1pt}-m'\hspace{1pt}q}^{j \hspace{6pt}j'\hspace{6pt}k}
\;\; (-1)^{l'-\lambda'}\; 
C_{\lambda\hspace{1pt}-\lambda'\hspace{1pt}q_l}^{l \hspace{6pt}l'\hspace{6pt}k_l} 
\;\; C_{q_l\hspace{1pt}q_s\hspace{1pt}q}^{k_l \hspace{1pt}k_s\hspace{1pt}k}
\left\{\begin{array}{ccc}  ~l~ ~l'~ ~k_l~ \\ ~s~ ~s~ ~k_s~ \\ ~j~ ~j'~ ~k~ 
           \end{array} \right\} \cr
&& \times Y_l^\lambda (1)\; R_{nlj}(1) \; Y_{l'}^{*\lambda'}(1') \; R^*_{n'l'j'}(1'),
\label{eq:2.5e}
\end{eqnarray}  
where all constants are collected in the number ${\mathcal N}$ and only the newly
introduced quantum numbers are shown in the sum.
From this expression, the terms related to the orbital angular momentum can be
extracted as
\begin{equation}
{\mathcal L}^{ll'}_{k_l q_l} (1,1') \equiv \sum_{l l'} (-1)^{l'-\lambda'} \; 
C_{\lambda\hspace{1pt}-\lambda'\hspace{1pt}q_l}^{l \hspace{6pt}l'\hspace{6pt}k_l} \;
Y_l^\lambda (1) \;\; Y_{l'}^{*\lambda'}(1').
\label{eq:2.5f}
\end{equation}
For evaluating the matrix element $\langle \phi_{I,M_I} |(a_{nljm})^\dagger
a_{n'l'j'm'} |\phi_{I,M'_I} \rangle$ let us consider 
\begin{eqnarray}
{\mathcal Q}_{k,q}& \equiv & \left\langle \phi_{I,M_I} \left|
\sum_{mm'} (-1)^{j'-m'} C_{m\hspace{1pt}-m'\hspace{1pt}q}^{j \hspace{6pt}j'\hspace{6pt}k} 
(a_{nljm})^\dagger a_{n'l'j'm'} \right|\phi_{I,M'_I} \right\rangle \cr
&=& C_{M'_I\hspace{1pt}q\hspace{1pt}M_I}^{I \hspace{6pt}k\hspace{6pt}I}\; \langle
\phi_{I,M_I} || \rho_k(nlj;n'l'j)|| \phi_{I,M'_I} \rangle ,
\label{eq:2.5g}
\end{eqnarray}
where the reduced matrix element consists of complex numbers and is independent of $M_I$,
$q$, and $ M'_I$. 

Thus, the angular momentum and spin structure of the single particle density matrix
is schematically given as
\begin{eqnarray}
\rho_{I,M_I;I,M'_I} (1,1') &\simeq& \sum_{k_l,q_l,k_s,q_s,k,q, \cdots} 
{\mathcal N} \; {\mathcal Q}_{q,k} \; {\mathcal L}^{l l'}_{k_l q_l} (1,1') \;
R_{nlj}(1) \; R^*_{n'l'j'}(1') \cr
& & \langle s m_s| \tau_{k_s q_s}(s) |s m'_s\rangle \;
C_{q_l\hspace{1pt}q_s\hspace{1pt}q}^{k_l \hspace{1pt}k_s\hspace{1pt}k}
\left\{\begin{array}{ccc}  ~l~ ~l'~ ~k_l~ \\ ~s~ ~s~ ~k_s~ \\ ~j~ ~j'~ ~k~ 
           \end{array} \right\}.
\label{eq:2.5h}
\end{eqnarray}

For a spin-zero target, $I=M_I=M'_I=0$, the Clebsch-Gordan coefficient in Eq.~(\ref{eq:2.5g})
requires $k=q=0$. Consequently, the Clebsch-Gordan coefficient of
Eq.~(\ref{eq:2.5h}) requires $k_s=k_l$. Thus, for $l=0$ only $k_s=0$ is possible,
i.e. the s-shell can not have any spin-dependent contribution.

For the consideration of 
$^6$He   we make  the assumption 
of an occupied s-shell, the alpha core, and the valence neutrons occupying the p-shell. 
We approximate the density matrix by two
harmonic oscillator terms. The one-particle s-wave harmonic oscillator wave function is
given by
\begin{equation} 
\Phi^m_s (\vvm p) =  \left( \frac{4}{\sqrt{\pi \nu_s^3}} \right)^{1/2}
e^{-p^2/2\nu_s} \; {\cal Y}_0^{\frac{1}{2},m}({\hat {\vvm p}}) \equiv f_s(p) \; {\cal
Y}_0^{\frac{1}{2},m}({\hat {\vvm p}}),
\label{eq:2.6}
\end{equation}
and the one-particle p-wave harmonic oscillator wave function by
\begin{equation} 
\Phi^m_p (\vvm p) =  \left( \frac{8}{3\sqrt{\pi \nu_p^{5}}}\right)^{1/2}
\;  p\; e^{-p^2/2\nu_p} \; {\cal Y}_1^{\frac{3}{2},m}({\hat {\vvm p}}) 
\equiv f_p(p)\; {\cal Y}_1^{\frac{3}{2},m} ({\hat {\vvm p}}).
\label{eq:2.7}
\end{equation}
Both wave functions are normalized to one.
The functions ${\cal Y}_l^{j=l\pm\frac{1}{2},m}({\hat {\bf p}})$ represent the total angular
momentum wave functions. The alpha-core consists of a filled s-shell contribution 
for protons as well as neutrons. According to  Eq.~(\ref{eq:2.5h})  the 
s-wave single-particle density matrix is a scalar function given by
\begin{equation}
\rho_s(\vvm p, \vvm p') = \Phi_s^* (\vvm p) \Phi_s (\vvm p') =
 \left( \frac{1}{\pi \nu_s}\right)^{\frac{3}{2}} e^{-\frac{p^2+p'^2}{2\nu_s}},
\label{eq:2.8}
\end{equation}
where the sum over $m$ has been carried out.

For the p-shell we make the assumption that the valence neutrons occupy the lowest possible
state, the $p_{3/2}$-shell. According to  Eq.~(\ref{eq:2.5h}), $k_l=1$, and both,
$k_s=0$ and $k_s=1$ are possible.   
%\begin{eqnarray}
%\rho^{tot}_p(\vvm p, \vvm p') =
%\frac{8}{3}\frac{pp'}{\sqrt\pi\nu^{5/2}}
%e^{-\frac{p'^2+p^2}{2\nu_p}} \; \sum_{m=-3/2}^{3/2}\left(
%C_{m\hspace{2pt}m\hspace{2pt}0}^{\frac{3}{2}\hspace{3pt}\frac{3}{2}\hspace{3pt}0}\right)^2
%\:\mathcal{Y}_{1}^{m*}(\hat{\bf{p}})\mathcal{Y}_{1}^{m}(\hat{\bf{p}}') 
%\label{eq:2.9}
%\end{eqnarray}
Evaluating the $k_s=0$ part for $l=l'=1$ according to Eq.~(\ref{eq:2.5h})
leads to
\begin{equation}
\rho_p(\vvm p, \vvm p')= \frac{2}{3} \left(\frac{1}{\pi^3 \nu_p^5} \right)^{\frac{1}{2}} \vvm p \cdot \vvm p' \;
e^{-\frac{p'^2+p^2}{2\nu_p}}.
\label{eq:2.9b}
\end{equation}
The contribution according to $k_s=1$ leads to a spin-dependent piece, which will
enter in the explicit calculation of the expectation values of spin-momentum
operators in Section~\ref{subvalence} and Appendix~\ref{appendixA}.

\begin{figure}
\begin{center}
\includegraphics[scale=.45]{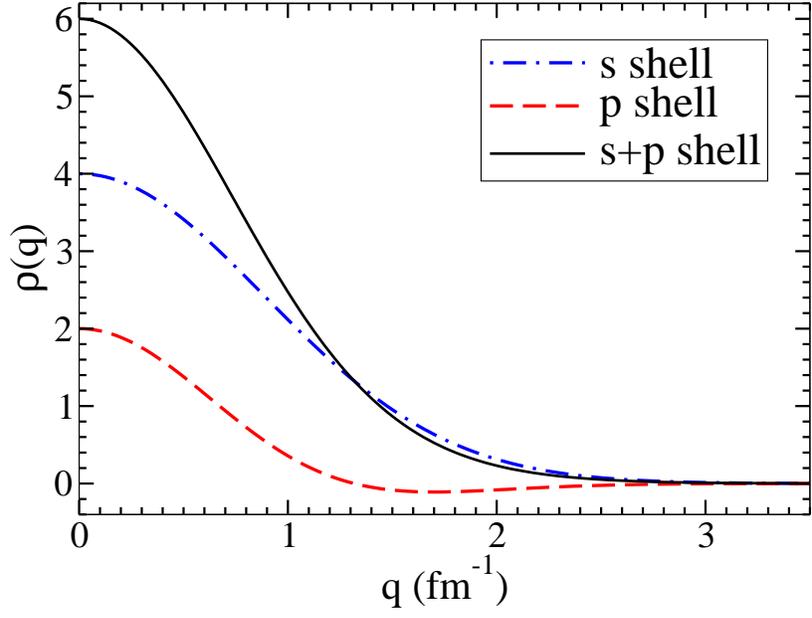}
\caption{(Color online) The diagonal density of $^6$He normalized to the
total particle number. The s-shell (dashed) and p-shell (dash-dotted) are given
separately. 
\label{fig2}
}
\end{center}
\end{figure}

\noindent
Changing variables in Eq.~(\ref{eq:2.9b}) to
\begin{eqnarray}
\vvm q &=& \frac{A}{A-1} (\vvm p - \vvm p') \cr
\vvm P &=& \frac{1}{2} (\vvm p + \vvm p')
\label{eq:2.10}
\end{eqnarray}
results in
\begin{eqnarray}
\vvm p \cdot \vvm p' &=& P^2-\left(\frac{A-1}{2A}\right)^2q^2 \cr
 p^2 + p'^2 &=& 2P^2+2\left(\frac{A-1}{2A}\right)^2  q^2.
\label{eq:2.11}
\end{eqnarray}
With these variables
the single-particle density matrices of Eqs.~(\ref{eq:2.8}) and (\ref{eq:2.9b}) become 
\begin{eqnarray}
\rho_s(\vvm q, \vvm P) &=& \left(\frac{1}{\pi\nu_s}\right)^\frac{3}{2} 
   e^{ -\frac{1}{\nu_s}\left( {P^2}+ \left(\frac{A-1}{2A}\right)^2 q^2 \right)} \cr
\rho_p(\vvm q, \vvm P) &=& \frac{2}{3} \left(\frac{1}{\pi^3\nu^5}\right)^{\frac{1}{2}} 
  \left( P^2 - \left(\frac{A-1}{2A}\right)^2 q^2\right) 
 e^{ -\frac{1}{\nu_p}\left( P^2 + \left(\frac{A-1}{2A}\right)^2 q^2\right)}.
\label{eq:2.12}
\end{eqnarray} 
From this we obtain the spin-independent single-particle density matrix of $^6$He as 
\begin{equation}
\rho_{^6\rm{He}} (\vvm q, \vvm P) = 4 \rho_s(\vvm q, \vvm P) + 2 \rho_p(\vvm q, \vvm P).
\label{eq:2.13}
\end{equation}
%and for $^8$He as
%\begin{equation}
%\rho_{^8\rm{He}} (\vvm q, \vvm P) = 4 \rho_s(\vvm q, \vvm P) + 4 \rho_p(\vvm q, \vvm P).
%\label{eq:2.14}
%\end{equation}
Integrating over the momentum $\vvm P$ leads to the diagonal density
\begin{eqnarray}
\rho_{^6\rm{He}} (\vvm q) = 
4 e^{-\left(\frac{A-1}{2A}\right)^2\:\frac{q^2}{\nu_s}}
 +2\left(1-\frac{q^2}{6\nu_p}\right) \; 
e^{-\left(\frac{A-1}{2A}\right)^2\:\frac{q^2}{\nu_p}}.
\label{eq:2.15}
\end{eqnarray}
It remains to determine the oscillator parameters for the two helium isotopes. The charge
radii for $^6$He~\cite{Mueller:2008bj} and $^8$He~\cite{Brodeur:2012zz} are very well
measured, and are used to determine the oscillator parameters for the s-shell according to
\begin{equation}
\langle r^2_{ch}\rangle = \frac{3}{2 \nu_s}.
\label{eq:2.16}
\end{equation}
The matter radius is determined by taking the expectation value of the radius with
 the total wave function. Using the prior determined s-shell oscillator parameter we obtain
the matter radius of $^6$He by
\begin{equation}
\langle r^2_{mat}\rangle = \frac{1}{6} \left( \frac{5}{\nu_p} + \frac{6}{\nu_s}\right)
\label{eq:2.17}
\end{equation}
and from this the value for $\nu_p$.
The experimental extractions of the matter radii used for our calculations are given in
Table~\ref{table-1}. The so obtained diagonal density for $^6$He is shown in Fig.~\ref{fig2} 
as function of the momentum transfer. The density is normalized such 
$\rho_{^6\rm{He}} (0) = 6$.

\subsection{Expectation Values of the Spin-Momentum Operators for the Target Nucleon}
\label{subvalence}

Having established a basis for the nuclear single-particle 
density matrix allows the calculation of
the matrix elements of the optical potential given in Eq.~(\ref{eq:2.4}). When
considering the first Wolfenstein amplitude in Eq.~(\ref{eq:2.5}), we encounter the unit
matrix between the plane wave and the nuclear basis states. This leads after a series of
variable transformations, which are in detail given in Ref.~\cite{Weppner:2011px}, 
to the central part of the optical potential 
\begin{eqnarray}
U_A (\vvm q, \vvm K )& =& \int d^3  P\;  A \left(\vvm q, \frac{1}{2}\left(\frac{A+1}{A}\vvm
K - \vvm P\right),{\cal{E}}\right) \rho_i\left(\vvm P-\frac{A-1}{2A}\vvm q, \vvm
P+\frac{A-1}{2A}\vvm q \right) \cr
 &=&  \int d^3  P\;  A \left(\vvm q, \frac{1}{2}\left(\frac{A+1}{A}\vvm
K - \vvm P\right),{\cal{E}}\right) \rho_{s(p)} (\vvm q, \vvm P),
\label{eq:2.2.1}
\end{eqnarray}
where $\vvm q$ is the momentum transfer, $\vvm K$ the momentum orthogonal to it, and
$\vvm P$ the  total momentum of the struck nucleon. 
The second line contains the explicit expressions for
the single-particle densities of Eq.~(\ref{eq:2.12}) and should be read as the sum over
the s- and p-shell contributions.

The next term in Eq.~(\ref{eq:2.5}) is proportional to 
$(\bsigma^{(0)} \otimes  I  + I \otimes \bsigma^{(i)})\cdot \hat {\bf n}_{NN}$, containing
the spin of the projectile as well as the spin of the struck nucleon tensorized with the
unit matrix in the respective space of the other nucleon. The term containing the spin
of the projectile leads to the well known spin-orbit term
\begin{eqnarray}
i \bsigma^{(0)} \cdot \hat {\bf n}_{NN}\;  U_C (\vvm q, \vvm K )& =& i \bsigma^{(0)} \cdot \hat
{\bf n}_{NN} \int d^3 P \; C \left(\vvm q, \frac{1}{2}\left(\frac{A+1}{A}\vvm
K - \vvm P\right),{\cal{E}}\right) \rho_i\left(\vvm P-\frac{A-1}{2A}\vvm q, \vvm
P+\frac{A-1}{2A}\vvm q \right) \cr
&=& i \bsigma^{(0)} \cdot \hat
{\bf n} \int d^3 P \; C \left(\vvm q, \frac{1}{2}\left(\frac{A+1}{A}\vvm
K - \vvm P\right),{\cal{E}}\right) \; \rho_{s(p)} (\vvm q, \vvm P) .
\label{eq:2.2.2}
\end{eqnarray}
All other terms in  Eq.~(\ref{eq:2.5}) contain the scalar products of the 
spin-operator of the struck nucleon with a momentum vector,
which needs to be evaluated in the nuclear intrinsic basis. For closed shell
nuclei, the sum over all possible magnetic quantum numbers of the total angular
momentum adds up to a zero contribution of those terms, as e.g. for $^{16}$O with a
filled s- and p-shell~\cite{Elster:1989en}. The alpha-core of $^6$He consists of a
filled s-shell, thus the optical potential for the s-shell only has a standard central
and spin-orbit term. For the p-shell, the considerations are more involved.

The evaluation of the spin-momentum operators for the target nucleon require several
steps. In principle they should be evaluated in the target intrinsic frame (TI), however the
NN t-matrix is given in its own NN frame.  For the momentum vectors given in the target
intrinsic frame we find for the expectation values of $\bsigma^{(i)}$ with the p$_{3/2}$
ground state wave function 
\begin{eqnarray}
\langle \Phi_p(\vvm p) | \bsigma^{(i)} \cdot \hat{\mathbf{q}}_{TI} | \Phi_p(\vvm p') \rangle 
&=& 0 \cr
\langle \Phi_p(\vvm p) | \bsigma^{(i)} \cdot \hat{\mathbf{P}}_{TI} | \Phi_p(\vvm p') \rangle &=& 0 \cr
\langle \Phi_p(\vvm p) | \bsigma^{(i)} \cdot \hat{\mathbf{n}}_{TI} | \Phi_p(\vvm p') \rangle &=&
-i \frac{2}{9} \frac{|\vvm p \times \vvm p'|}{\sqrt{\pi^3 \nu_p^5}}
\exp \left(-\frac{p^2+p'^2}{2\nu_p}\right).
\label{eq:2.2.3}
\end{eqnarray}
The momentum transfer $\vvm q$ has a special role, since it is invariant in all frames.
Thus the scalar product $(\bsigma^{(i)} \cdot \vvm q)$ will always give a zero
contribution. Next, the expectation values of Eq.~(\ref{eq:2.2.3}) needs to be projected
into the NN frame, where the Wolfenstein amplitudes are defined. The details are
given in Appendix~\ref{appendixA} and summarized as
\begin{eqnarray}
\langle \Phi_p(\vvm p) | \bsigma^{(i)} \cdot \hat{\mathbf{q}}_{NN}| \Phi_p(\vvm p') \rangle &=& 0 \cr
\langle \Phi_p(\vvm p) | \bsigma^{(i)} \cdot \hat{\mathbf{n}}_{NN}| \Phi_p(\vvm p') \rangle &=&
-i \frac{2}{9} \frac{|\vvm p \times \vvm p'|}{\sqrt{\pi^3 \nu_p^5}} \cos \beta \;
e^{-\frac{p^2+p'^2}{2}}\cr
\langle \Phi_p(\vvm p) | \bsigma^{(i)} \cdot \hat{\mathbf{K}}_{NN}| \Phi_p(\vvm p') \rangle &=&
-i \frac{2}{9} \frac{|\vvm p \times \vvm p'|}{\sqrt{\pi^3 \nu_p^5}} \cos \alpha \;
e^{-\frac{p^2+p'^2}{2}},
\label{eq:2.2.4}
\end{eqnarray}
where $\cos \beta = {\hat {\bf n}}_{TI}\cdot {\hat {\bf n}}_{NN}$ and 
$\cos \alpha = {\hat {\bf n}}_{TI}\cdot {\hat {\bf K}}_{NN}$. 

Considering the expression for the NN t-matrix of Eq.~(\ref{eq:2.5}), we note that 
terms that contain $(\bsigma^{(i)} \cdot \hat{\mathbf{q}})$ vanish. This corresponds to
the term proportional to ($G-H$) and one term proportional to $D$. 
The remaining terms will in principle all contribute to the optical potential for the
valence neutrons.

Let us first consider the term of the scattering amplitude, Eq.~(\ref{eq:2.5}), proportional to
$i C(I\otimes \bsigma^{(i)})\cdot \hat{\mathbf{n}}_{NN}$. Inserting the expectation value 
of Eq.~(\ref{eq:2.2.4}) and transforming to the variables {\bf q} and {\bf K} 
in the nucleon-nucleus frame leads to a term 
\begin{equation} 
i \; U_A^C(\vvm q, \vvm K) =i \int d^3 P \; C\left(\vvm q, \frac{1}{2}\left(\frac{A+1}{A}\vvm K 
- \vvm P\right),{\cal{E}}\right)  {\tilde \rho}_p  (\vvm q, \vvm P) \cos \beta,
\label{eq:2.2.5}
\end{equation}
with
\begin{equation}
{\tilde \rho}_p (\vvm q, \vvm P) = -i N_p \frac{2}{9} \frac{1}{\sqrt{\pi^3 \nu_p^5}} \;| \vvm q \times \vvm P| \;
      e^{-\frac{1}{\nu_p}\left(P^2 +\left(\frac{A-1}{2A}\right)^2 q^2\right)}.
\label{eq:2.2.6}
\end{equation}
Here $N_p$ denotes the number of valence neutrons in the p$_{3/2}$-shell.
The term of Eq.~(\ref{eq:2.2.5}) does not contain any spin dependence and 
thus contributes to the central part of the 
optical potential.  Comparing ${\tilde \rho}_p (\vvm q, \vvm P) $ with the p-shell
single-particle density matrix of Eq.~(\ref{eq:2.12}) reveals, that ${\tilde \rho}_p
(\vvm q, \vvm P) $ is  reduced by a factor of three
 and contains the cross product  $\vvm q \times \vvm P$.
The latter corresponds to the structure expected from Eq.~(\ref{eq:2.5f}) for $k_l=1$. 

The Wolfenstein amplitude $M$ is proportional to $(\bsigma^{(0)} \cdot \hat{\mathbf{n}}_{NN}) \otimes
(\bsigma^{(i)} \cdot \hat{\mathbf{n}}_{NN})$, and thus leads to the same expectation
value ${\tilde \rho}_p
(\vvm q, \vvm P)$ when evaluated for the struck nucleon,
\begin{equation}
 \bsigma^{(0)} \cdot \hat {\bf n}_{NN}\; U^M (\vvm q, \vvm K) =\bsigma^{(0)} \cdot \hat
{\bf n}_{NN}
\int d^3 P \;M\left(\vvm q,\:\frac{1}{2}\left(\frac{A+1}{A}\vvm K-\vvm P\right),\mathcal
E\right) {\tilde \rho}_p (\vvm q,\vvm P) \cos \beta .
\label{eq:2.2.7}
\end{equation} 

The remaining non-vanishing terms of ${\overline M}$ in Eq.~(\ref{eq:2.5}) have a slightly
different character, they are proportional to $(\bsigma^{(0)} \cdot
\hat{\mathbf{K}}_{NN})$ and $(\bsigma^{(0)} \cdot \hat{\mathbf{q}}_{NN})$ as far as the
projectile is concerned.  These scalar products need to be projected on spin-flip and
non-spin-flip amplitudes in order to classify them as terms which contribute to the
central (non-spin-flip) and to the spin-orbit (spin-flip) terms in the optical
potential for scattering of a spin-0 from a spin-1/2 particle. 
The projection of the Wolfenstein amplitude $(G+H)$ on the central and spin-orbit term
leads to 
\begin{eqnarray}
U_A^{G+H}(\vvm q, \vvm K) &=& \int d^3 P  \; \left\{ G \left(\vvm q,\frac{1}{2}
\left(\frac{A+1}{A}\vvm K-\vvm P\right),{\mathcal E}\right)+
H\left(\vvm q,\frac{1}{2}\left(\frac{A+1}{A}\vvm K-\vvm P\right),{\mathcal
E}\right)\right\} \cr
 &&\frac{1}{2|K_{NN}|} \left(|k_{NN}|+|k_{NN}'|\cos\gamma_{NN}\right)
\:\cos\alpha \; {\tilde \rho} (\vvm q, \vvm P)  \cr
\bsigma^{(0)}\cdot\hat{\vvm n}_{NN} \;  U_C^{G+H}(\vvm q,\vvm K)&=&
\bsigma^{(0)}\cdot\hat{\vvm n}_{NN} \int d^3 P \; 
\left\{ G \left(\vvm q,\frac{1}{2} \left(\frac{A+1}{A}\vvm K-\vvm P\right),{\mathcal E}\right)+
H\left(\vvm q,\frac{1}{2}\left(\frac{A+1}{A}\vvm K-\vvm P\right),{\mathcal
E}\right)\right\} \cr 
&& \frac{(-i)}{2|K_{NN}|} |k_{NN}'|\sin\gamma_{NN}\; \cos \alpha \;\tilde\rho(\vvm q,\vvm
P)
\label{eq:2.2.8}
\end{eqnarray} 
Here the angle $\gamma_{NN}$ is the angle between the momenta ${\bf k}_{NN}$ and ${\bf
k'}_{NN}$ in the NN frame. The vector ${\bf K}_{NN}$ is defined in the same way as the
vector {\bf P} of Eq.~(\ref{eq:2.10}).

The non-vanishing term of the Wolfenstein amplitude $D$ leads to 
\begin{eqnarray}
U_A^{D}(\vvm q, \vvm K) &=& \int d^3 P \;  D\left(\vvm q,\frac{1}{2}
\left(\frac{A+1}{A}\vvm K-\vvm P\right),{\mathcal E}\right)
\;\frac{1}{|q|}\;(|k_{NN}'|\cos\gamma_{NN}-|k_{NN}|)\:\cos\alpha \;\tilde\rho(\vvm q,\vvm
P) \cr
\bsigma^{(0)}\cdot\hat{\vvm n}_{NN} \;  U_C^D (\vvm q,\vvm K)&=& 
\bsigma^{(0)}\cdot\hat{\vvm n}_{NN} \int d^3 P \;  D \left(\vvm q,\frac{1}{2}
\left(\frac{A+1}{A}\vvm K-\vvm P\right),{\mathcal E}\right)
\;\frac{(-i)}{|q|}\;|k_{NN}'|\sin\gamma_{NN}\:\cos\alpha \;\tilde\rho(\vvm q,\vvm P).
\label{eq:2.2.9}
\end{eqnarray}
The explicit calculation of the integrals of Eqs.~(\ref{eq:2.2.8}) and (\ref{eq:2.2.9}) 
reveals that the contributions of $U^{G+H}$ and $U^D$ vanish since the integrands of
Eqs.~(\ref{eq:2.2.8}) and (\ref{eq:2.2.9}) are odd functions of one of the integration angles. 
Elements of the explicit
proof of this result are given in Appendix~\ref{appendixB}. The physical interpretation
of this result may stem from the fact that the amplitudes $G$, $H$, and $D$ are related to
the NN tensor force. 
Since we work with one oscillator wave function in the p-shell, we have $l=l'=1$ in
Eq.~(\ref{eq:2.5f}), which excludes contributions of the tensor force.

\subsection{Elastic Scattering Observables for $^6$He} 
\label{subobshe6}

In Section~\ref{subdenshe6} we derived a model single-particle density for the $^6$He nucleus consisting of a
filled s-shell, the alpha-core, and two valence neutrons in the p$_{3/2}$ sub-shell, coupled to a total spin
zero. 
In this case, the contributions proportional to the Wolfenstein amplitudes ($G+H)$ and $D$ vanish,
leading to an optical potential of the form
\begin{equation}
U(\vvm q, \vvm K) = U_A(\vvm q, \vvm K) + i U^C_A(\vvm q, \vvm K) + i\; \bsigma^{(0)}\cdot {\hat {\vvm n}}
\left\{ U_C(\vvm q, \vvm K) -i\; U^M(\vvm q, \vvm K)\right\}.
\label{eq:2.3.1}
\end{equation}
The terms $U_A(\vvm q, \vvm K)$ and $U_C(\vvm q, \vvm K)$ contain the contributions from the s- as well
as the p-shell and have been traditionally calculated for microscopic optical potentials for 
closed shell nuclei. The terms  $U^C_A(\vvm q, \vvm K)$ and $U^M(\vvm q, \vvm K)$ result from 
the explicit evaluation of spin-momentum operators of the struck target nucleons in the p$_{3/2}$
sub-shell.

 The oscillator parameters of the single-particle nuclear density matrix
 are fitted to the charge radius~\cite{Mueller:2008bj} and the matter
radius~\cite{wang-thesis} of $^6$He. For this specific ground state configuration we calculate the
additional terms that arise from explicitly evaluating the expectation values of
the spin-momentum operators of the struck target nucleon with these ground state wave functions.
We find that this particular choice of ground state wave functions leads to two additional terms in 
the optical potential, one that is spin independent and proportional to the Wolfenstein amplitude $C$, adding
to the central part of the optical potential, and one spin dependent term proportional to the Wolfenstein
amplitude $M$ adding to the spin-orbit part.  

\begin{figure}[htbp]
\begin{center}
\includegraphics[scale=.75]{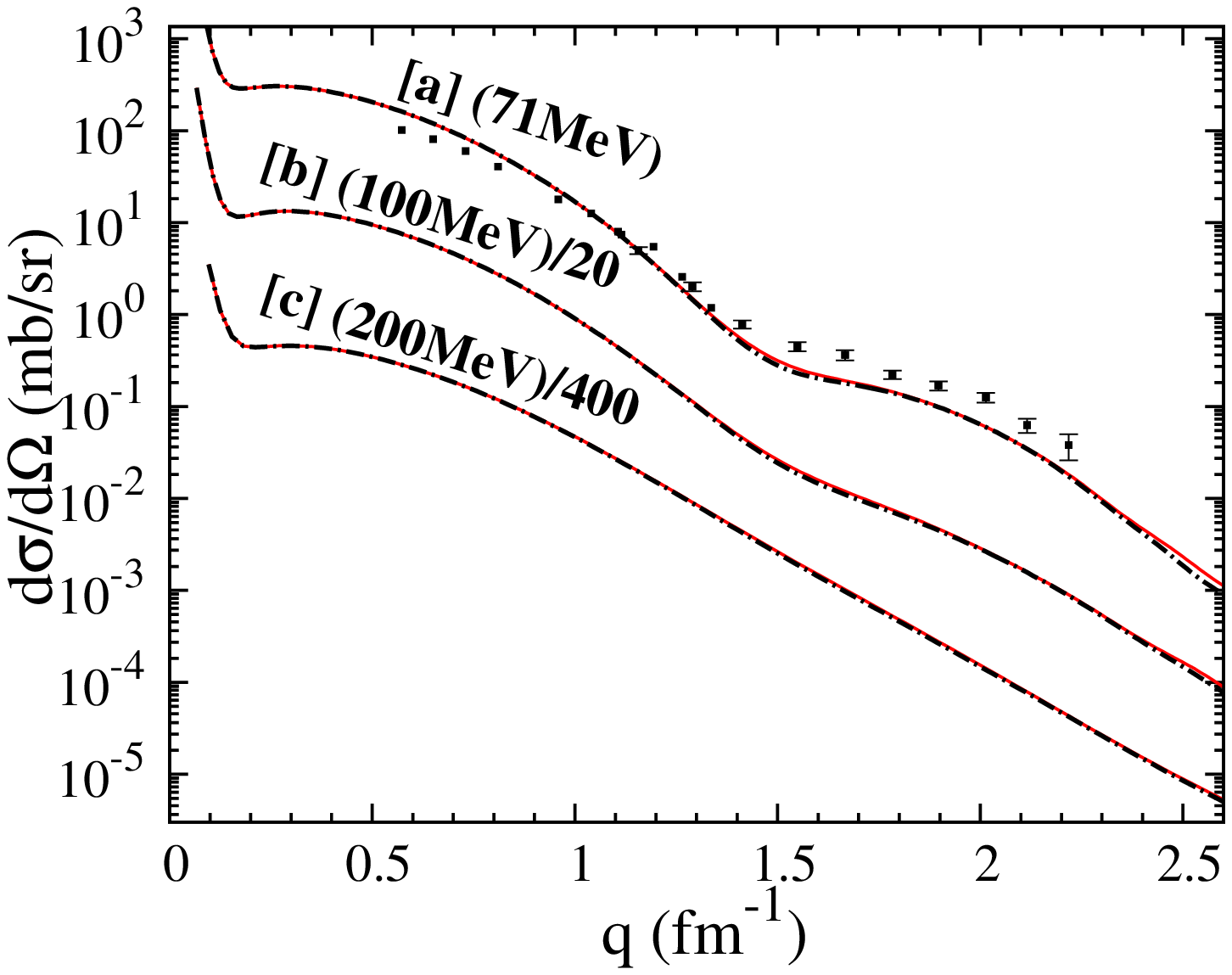}
\caption{(Color online) The angular distribution of the differential cross
section, $\frac{d\sigma}{d\Omega}$, for elastic scattering of $^{6}$He at
projectile energies [a] 71~MeV/nucleon, [b] 100~MeV/nucleon/40, and [c]
200~MeV/nucleon/400 as function of the momentum transfer. The calculations are
performed with an optical potential based on the CD-Bonn potential. The solid line
represents the full calculations, while the dash-dotted line represents the
calculating omitting open-shell effects. The data are taken from
Refs.~\cite{Uesaka:2010mm,Korsheninnikov:1997mm}
\label{fig3}
}
\end{center}
\end{figure}

\begin{figure}[htbp]
\begin{center}
\includegraphics[scale=.75]{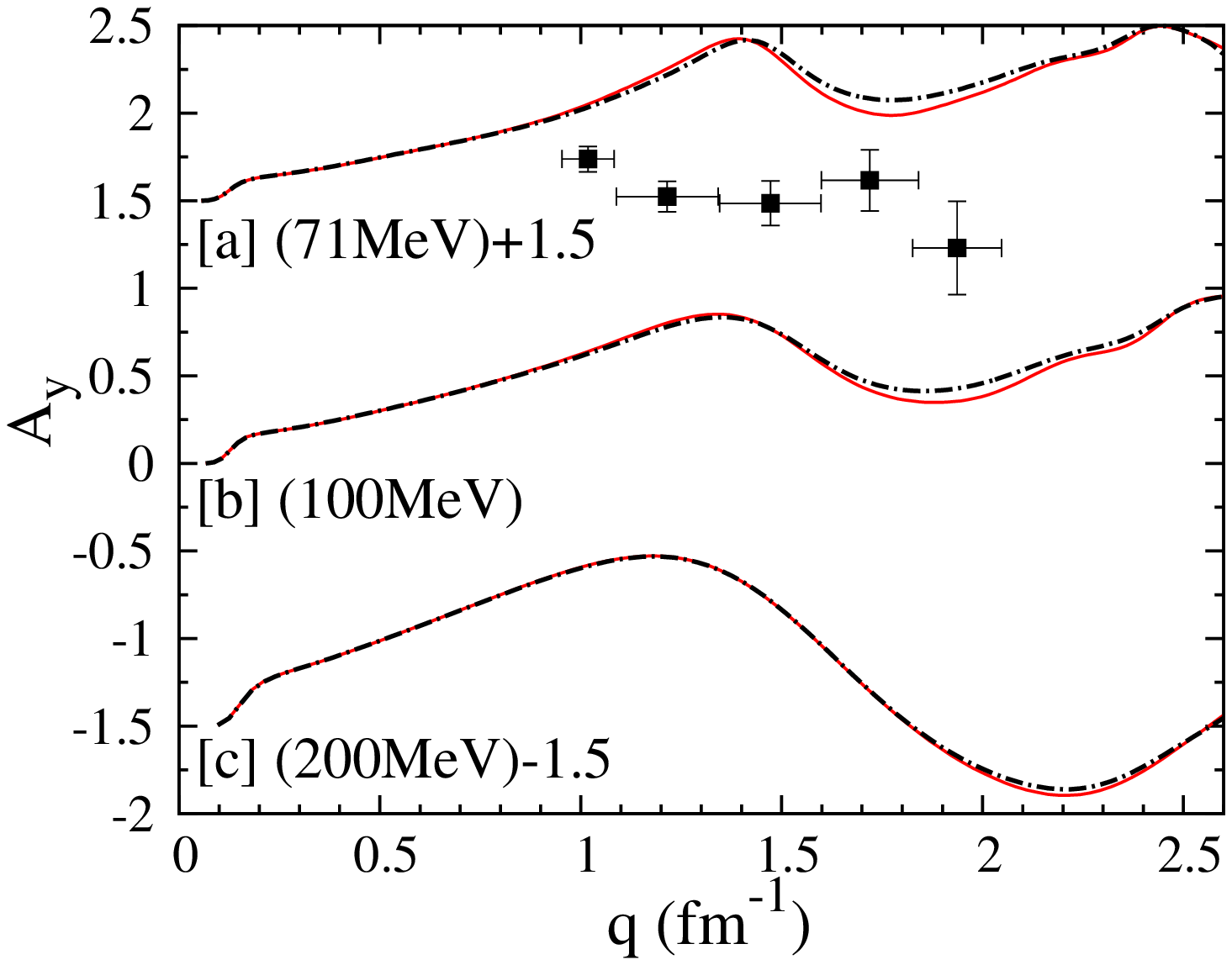}
\caption{(Color online) The angular distribution of the analyzing power ($A_y$)
for elastic scattering of $^{6}$He at
projectile energies [a] 71~MeV/nucleon, [b] 100~MeV/nucleon, and [c]
200~MeV/nucleon as function of the momentum transfer. The values for [a] and [c]
are shifted as indicated in the figure.  The meaning of
the lines is the same as in Fig.~\ref{fig3}. The data are taken from
Ref.~\cite{Uesaka:2010mm,Sakaguchi:2011rp}.
\label{fig4}
}
\end{center}
\end{figure}

In order to study the effect of those two additional term we first calculate the
differential cross section, $d\sigma/d\Omega$,  and the analyzing power, $A_y$, 
for scattering of $^6$He from a polarized proton target using a
a folding optical potential based only on the traditionally used central and spin-orbit terms corresponding
to the Wolfenstein amplitudes $A$ and $C$. Those calculations are shown by the dashed lines in
Fig.~\ref{fig3} for the differential cross section and Fig.~\ref{fig4} for the analyzing power. Our
calculations are carried out for 71, 100 and 200~MeV per nucleon, and use the CD-Bonn
potential~\cite{Machleidt:2000ge} as NN interaction. Then we add the two additional contributions from the
valence neutrons to the optical potential and show those calculations as solid lines in Figs.~\ref{fig3} and
\ref{fig4}. First we notice, that the differential cross section is completely insensitive the
additional terms. This might be expected since the expectation value ${\tilde \rho}$ is an order of
magnitude smaller than the single-particle density matrix. However, the effect of additional contribution to
the spin-orbit potential through the Wolfenstein amplitude $M$ is also very small. We note, that 
there is also a small effect on $A_y$ through the change in the central potential. However, both effects are
so small, that they do not warrant to be shown separately. 

In closing this section, we want to comment on final state interactions resulting
from the breakup of the $^6$He on the scattering
process. The effect of final state interactions in a proton-nucleus optical potential was
studied in Ref.~\cite{Elster:1997as} for closed shell nuclei, with $^{16}$O being the
lightest nucleus, 
for projectile energies between 65 and 200~MeV. This study concluded that for
projectile energies of 100~MeV and above there was no effect, and at 65~MeV it was very
small. We expect that this conclusion will also hold in the case of $^6$He 
scattering off a
proton target, since in this case the breakup of $^6$He would lead to a 
$np$ final state interaction, which is strongest when the $np$ system is in an s-wave and
the relative energy of the $np$-pair is less than 10~MeV. Even the lowest energy we
consider, namely 71~MeV, is sufficiently high, that we are quite certain that $np$ final
state interactions are too small to affect the results of our calculations.

%%%%%%%%%%%%%%%%%%%%%%%%%%%%%%%%%%%%%%%%%%%%%%%%%%%%%%%%%%%%%%%%%%%%%%%%%%

\section{Sensitivity of the $^{6}$He Scattering Observables to the Functional 
Form of the Wave Function for large Radii}
\label{exptail}

In the previous section we calculated additional contributions to the optical potential
for $^6$He due to the two valence neutrons occupying the p$_{3/2}$ ground state, and 
find that their effect on the observables for elastic scattering is very small. 
We use a very simple ansatz for the single-particle density matrix, namely only two
harmonic oscillator functions, which may lead to this very small contribution.
A further point of concern is the asymptotic behavior of the harmonic oscillator
wave functions, which do not correctly capture the halo character of the $^6$He nucleus. 
Therefore, we need to investigate, if the behavior of the wave functions for large values of
$r$, i.e. the tail of the coordinate space wave function, can be seen in the scattering
observables at the energies we consider.  For the calculation of S-factors, i.e. at very
low energies, it is well
known that the asymptotic form of the nuclear wave functions is very
important~\cite{Navratil:2006tt}. We need to carry out a similar investigation for
our calculations. 

Considerations about the asymptotic behavior of the single-particle wave functions are
most naturally carried out in coordinate space, though we will have to define some
`equivalent' in momentum space. Following a similar line of thought as
Ref.~\cite{Navratil:2006tt} we define the radial part of the p-shell wave function as
\begin{equation}
\Phi^{radial}_p (r) = \left( 2\sqrt{\frac{1}{6}}\;\frac{r \;\nu_p^{5/4}}{\pi^{1/4}} \;
 e^{\left(-\frac{r^2\;\nu_p}{2}\right)} \bigg|_{r\leq R_{m}} +
 B\;e^{-\mu\;r}\bigg|_{r>R_{m}}\right)
%\left(\mathcal{Y}_1^{3/2}({\hat r})-\mathcal{Y}_1^{1/2}({\hat r})
%+\mathcal{Y}_1^{-1/2}({\hat r})-\mathcal{Y}_1^{-3/2}({\hat r})\right).
\label{eq:3.1}
\end{equation}  
Here $R_m$ is the matching radius, at which we match the 
harmonic oscillator p-wave and its derivative with an exponential tail. The parameter 
$\mu$ should in principle be close to the two-nucleon separation energy of the valence
neutrons. The oscillator parameters are $\nu_s = 0.392$~fm$^{-1}$ and $\nu_p =
0.289$~fm$^{-1}$. 

\begin{figure}[htbp]
\begin{center}
\includegraphics[scale=.45]{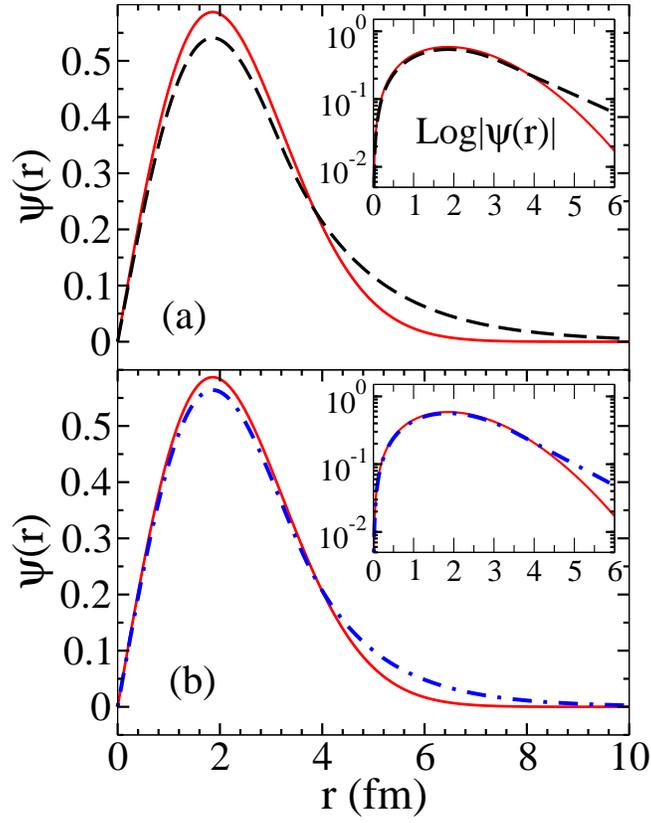}
\caption{(Color online) The coordinate space p-shell wave function of $^6$He. The
solid (red) line represents a harmonic oscillator wave functions.  For the dashed
(black) line in panel (a) an exponential functions was matched at $R_m$~=~3.2~fm,
while for the dash-dotted (blue) line in panel (b) an exponential function was
matched at $R_m$~=~3.5~fm.
\label{fig5}
}
\end{center}
\end{figure}

For determining reasonable values for $R_m$ we want to assume, that the alpha-core of
$^6$He shall not be significantly affected by changing the behavior of the p-wave. Thus we
ensure that for fixed $R_m$ the integral over the  s-wave harmonic
oscillator function
contains most of the mass of the alpha core. The s-shell probability
is given in  Table~\ref{table-2} as function of $R_m$. The values for the s-shell probability 
show, that for $R_m \geq 3.2$~fm more than 95\% of the alpha core are being described by
the s-wave oscillator function, and thus the core is minimally affected by the matching
procedure. For $R_m = 3.2$~fm about 69\% of the probability for the valence neutrons is
described by the p-shell oscillator wave function, the remaining by the exponential
tail. Normalizing this hybrid p-wave leads to a norm of 2.35, and we have to renormalize 
the p-wave to two, the number of neutrons in the p-shell. Choosing $R_m = 3.5$~fm leaves
almost the entire alpha-core unmodified, describes about 79\% of the valence
neutrons by the harmonic oscillator p-wave, and gives a norm of 2.17.
Table~\ref{table-2} shows in addition those values for $R_m = 2.8$~fm and $R_m = 3.8$~fm.
The small value of $R_m$ gives a p-shell norm of three, which means that there
would be three valence neutrons. For this reason we consider $R_m = 2.8$~fm as too small a
matching radius. 
The highest value in Table~\ref{table-2} is  $R_m = 3.8$~fm, which we consider a bit large
for studying effects of a change in the p-wave tail. We included the values in the
table, to support our arguments for choosing $R_m = 3.2$~fm and $R_m = 3.5$~fm for our study
of the sensitivity to an exponential tail of the p-wave on the scattering observables.
The coordinate space p-shell wave functions for those two cases are shown in 
Fig.~\ref{fig5} in comparison with the original harmonic oscillator p-wave. 
For completeness Table~\ref{table-2} also contains the matter radii $r_{mat}$ calculated
with the modified p-waves. Once the parameters $\mu$ and $B$ for the exponential tail
are determined through matching the logarithmic derivative at $R_m$ and renormalizing
the p-wave probability to two neutrons, the matter radius is a predictive quantity.

\begin{figure}[htbp]
\begin{center}
\includegraphics[scale=.35]{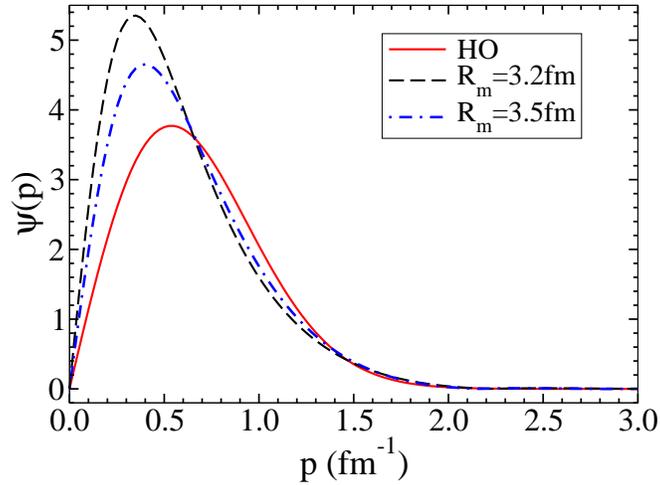}
\caption{(Color online) The momentum space p-shell wave function of $^6$He.
The solid (red) line represents a harmonic oscillator wave function, while for the
dashed (black) line an exponential function was matched at $R_m$~=~3.2~fm, and for
the dash-dotted (blue) line an exponential function was matched at $R_m$~=~3.5~fm.
\label{fig6}
}
\end{center}
\end{figure}

For the momentum space calculations we need to Fourier transform the wave functions 
and renormalize them to the number of nucleons in $^6$He. The resulting momentum space
p-waves are shown in Fig.~\ref{fig6} together with the original harmonic oscillator
p-wave. This figure also indicates, that an exponential
tail in coordinate space leads to a modification of the momentum space
wave function for small momenta.  From these wave functions we construct the
single-particle density matrix and calculate the microscopic folding optical potential. 

\begin{figure}[htbp]
\begin{center}
\includegraphics[scale=.75]{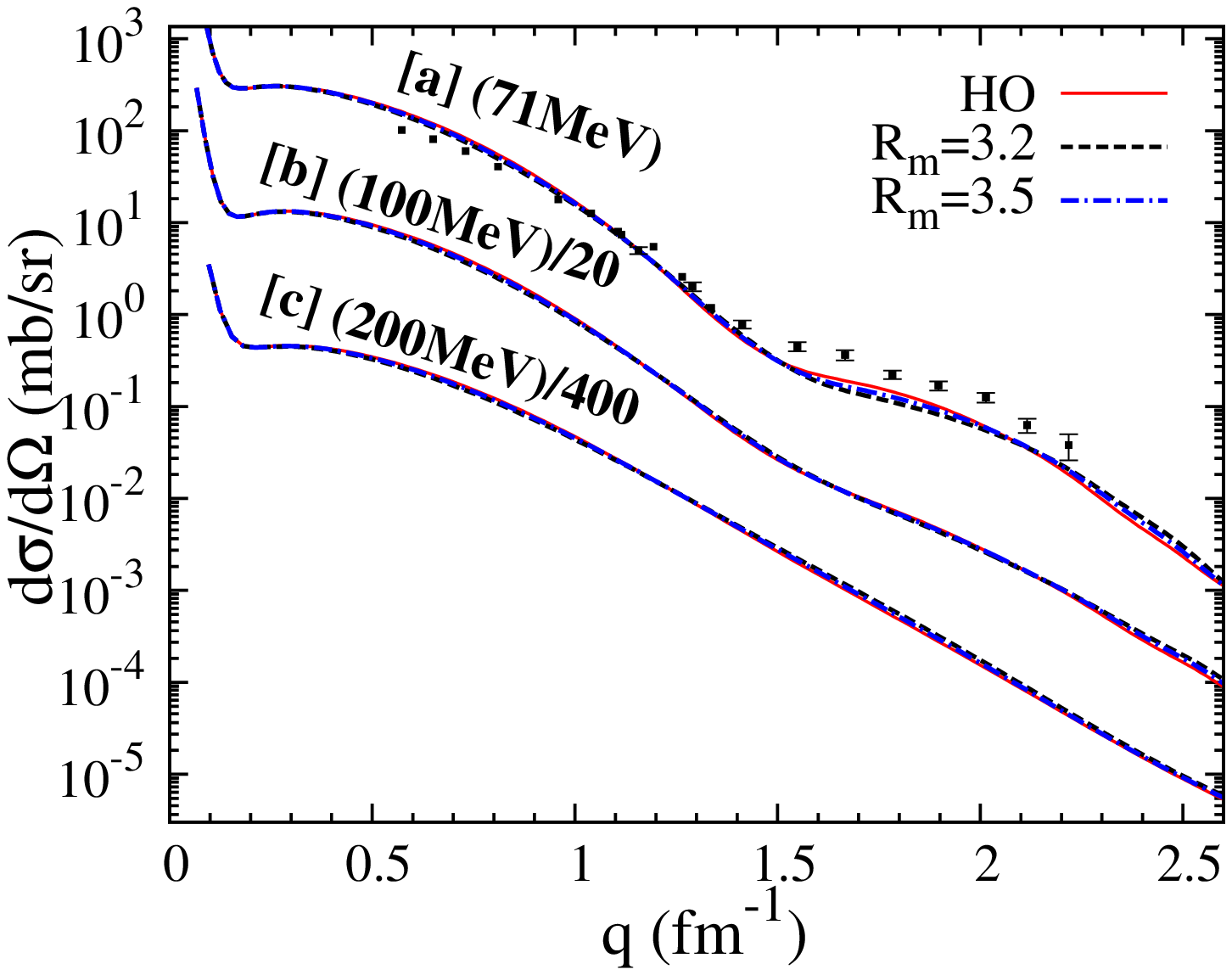}
\caption{(Color online) The angular distribution of the differential cross
section, $\frac{d\sigma}{d\Omega}$, for elastic scattering of $^{6}$He at
projectile energies [a] 71~MeV/nucleon, [b] 100~MeV/nucleon/40, and [c]
200~MeV/nucleon/400 as function of the momentum transfer. The calculations are
performed with an optical potential based on the CD-Bonn potential. The solid
(red) line represents the full calculations based on Harmonic Oscillator densities, 
while the dashed (black) line represents a calculation in which the p-shell
neutron wave functions have been matched at $R_m$~=~3.2~fm with an exponential
wave function. For the dash-dotted (blue) line this matching radius is
$R_m$~=~3.m~fm.
The data are taken from
Refs.~\cite{Uesaka:2010mm,Korsheninnikov:1997mm}
\label{fig7}
}
\end{center}
\end{figure}

\begin{figure}[htbp]
\begin{center}
\includegraphics[scale=.75]{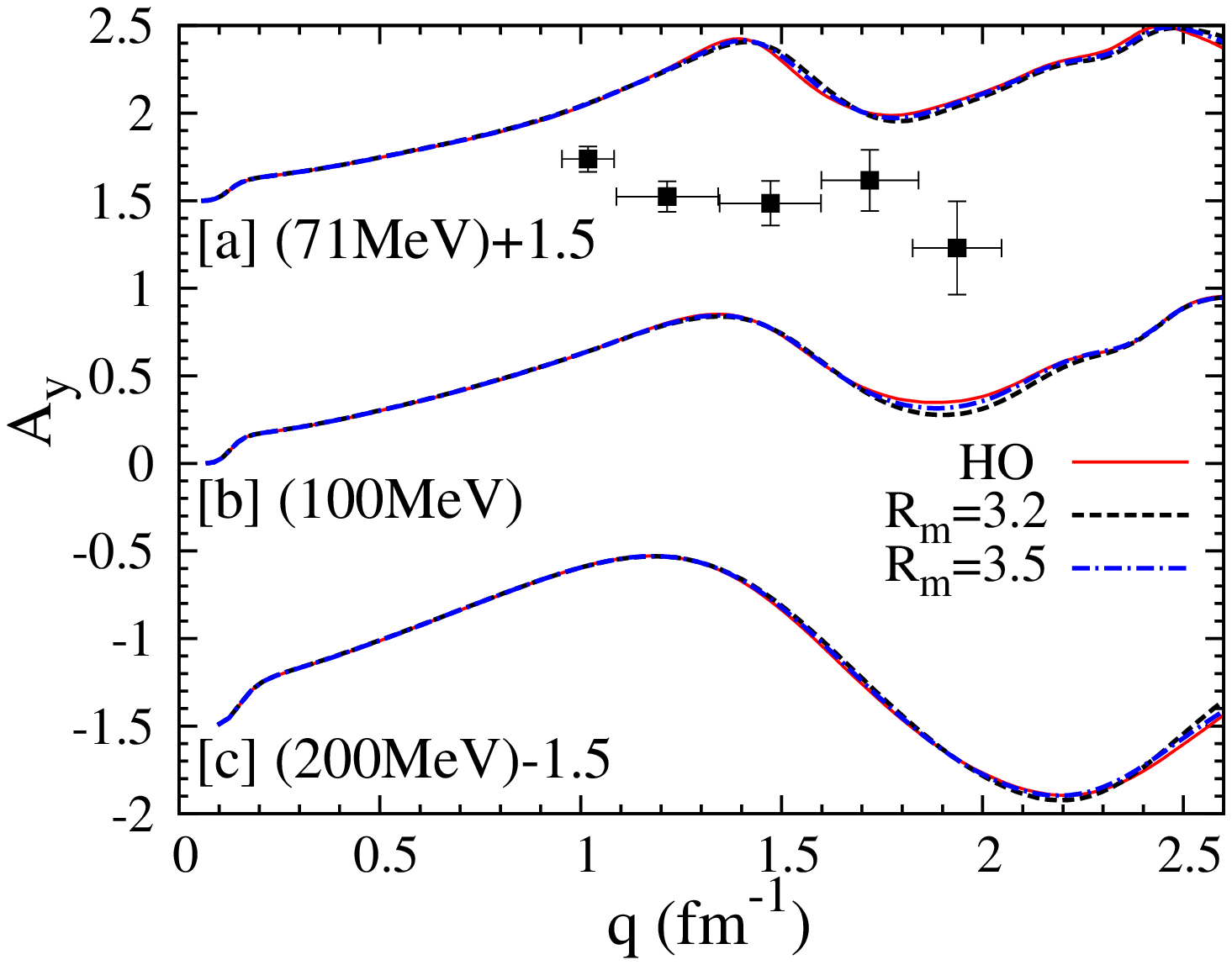}
\caption{(Color online) The angular distribution of the analyzing power ($A_y$)
for elastic scattering of $^{6}$He at
projectile energies [a] 71~MeV/nucleon, [b] 100~MeV/nucleon, and [c]
200~MeV/nucleon as function of the momentum transfer. The values for [a] and [c]
are shifted as indicated in the figure.  The meaning of
the lines is the same as in Fig.~\ref{fig7}. The data are taken from
Ref.~\cite{Uesaka:2010mm,Sakaguchi:2011rp}.
\label{fig8}
}
\end{center}
\end{figure}

The calculations of the differential cross section for 71, 100, and 200~MeV per nucleon
are shown in Fig.~\ref{fig7}. The figure shows that the different exponential tails of the
p-wave have no effect on this observable. In Fig.~\ref{fig8} we show the corresponding
calculations of the angular distribution of the analyzing power. Again, the
exponential form of the p-wave tail has no effect on this observable.  

The different functional of the tail of the coordinate space p-wave translates into
differences in the p-wave for small momentum p for the momentum space p-wave.       
Our calculations of the scattering observables for projectile energies from 71 to
200~MeV per nucleon show, that for these energies the affected small momenta of the
single-particle density have no effect on the observables. This conclusion is quite different
from the one in Ref.\cite{Navratil:2006tt} in which the extraction of S-factors from
reactions below 1~MeV was investigated. These two finding are not in contradiction, since at
very low energies, reactions are expected to be mostly sensitive to the long range part of
wave functions, whereas for the higher energy regime considered in this work, the asymptotic
part of the wave functions, and thus single particle density matrices should play a lesser
role.

\section{Sensitivity of the Scattering Observables to the Charge and Matter
Radii of $^6$He}

After establishing that at the scattering energies under consideration the fall-off behavior of
the wave functions in coordinate space has no significant effect on the scattering observables, we
should study if other input parameters into our model lead to discernible effects. In Section
\ref{subobshe6} we presented calculations for the differential cross section and the analyzing
power using oscillator parameters from Table~\ref{table-1}. Over the last years there have been
several measurements of the charge radius of $^6$He. Our model density uses the charge radius
to determine the oscillator parameter $\nu_s$ for the s-shell single particle density according
to Eq.~(\ref{eq:2.16}). The s-shell single particle density determines the size of the
alpha-core in our model, and therefore we want to test, how sensitive the elastic scattering
observables are to changes in $\langle r^2_{ch}\rangle$. As limits for this check we use the
measurement of Ref.~\cite{Wang:2004ze}, which obtained a charge radius of 1.894~fm as lower
limit and the value of 1.996~fm~\cite{Mueller:2008bj} as upper limit. 

\begin{figure}[htbp]
\begin{center}
\includegraphics[scale=.75]{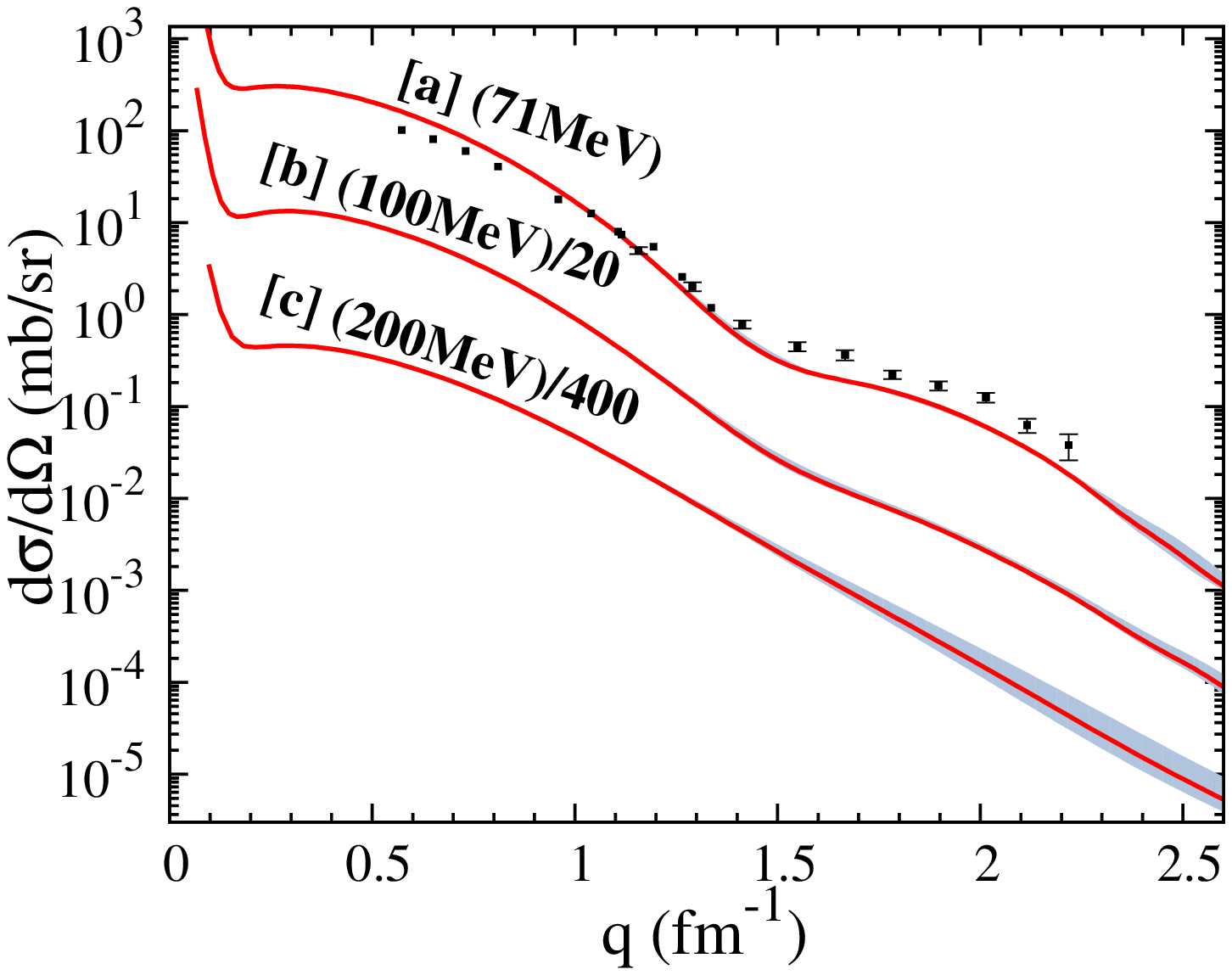}
\caption{(Color online) The angular distribution of the differential cross
section, $\frac{d\sigma}{d\Omega}$, for elastic scattering of $^{6}$He at
projectile energies [a] 71~MeV/nucleon, [b] 100~MeV/nucleon/40, and [c]
200~MeV/nucleon/400 as function of the momentum transfer. The solid (red) line
corresponds to the same calculation as the solid line in Fig.~\ref{fig3}, which 
which gives a charge radius $r_{ch}$~=~1.955~fm and a matter radius
$r_{mat}$~=~2.33~fm. The shaded region shows the variation of the charge radius
from 1.89~fm to 1.99~fm. 
The data are taken from
Refs.~\cite{Uesaka:2010mm,Korsheninnikov:1997mm}
\label{fig9}
}
\end{center}
\end{figure}

\begin{figure}[htbp]
\begin{center}
\includegraphics[scale=.75]{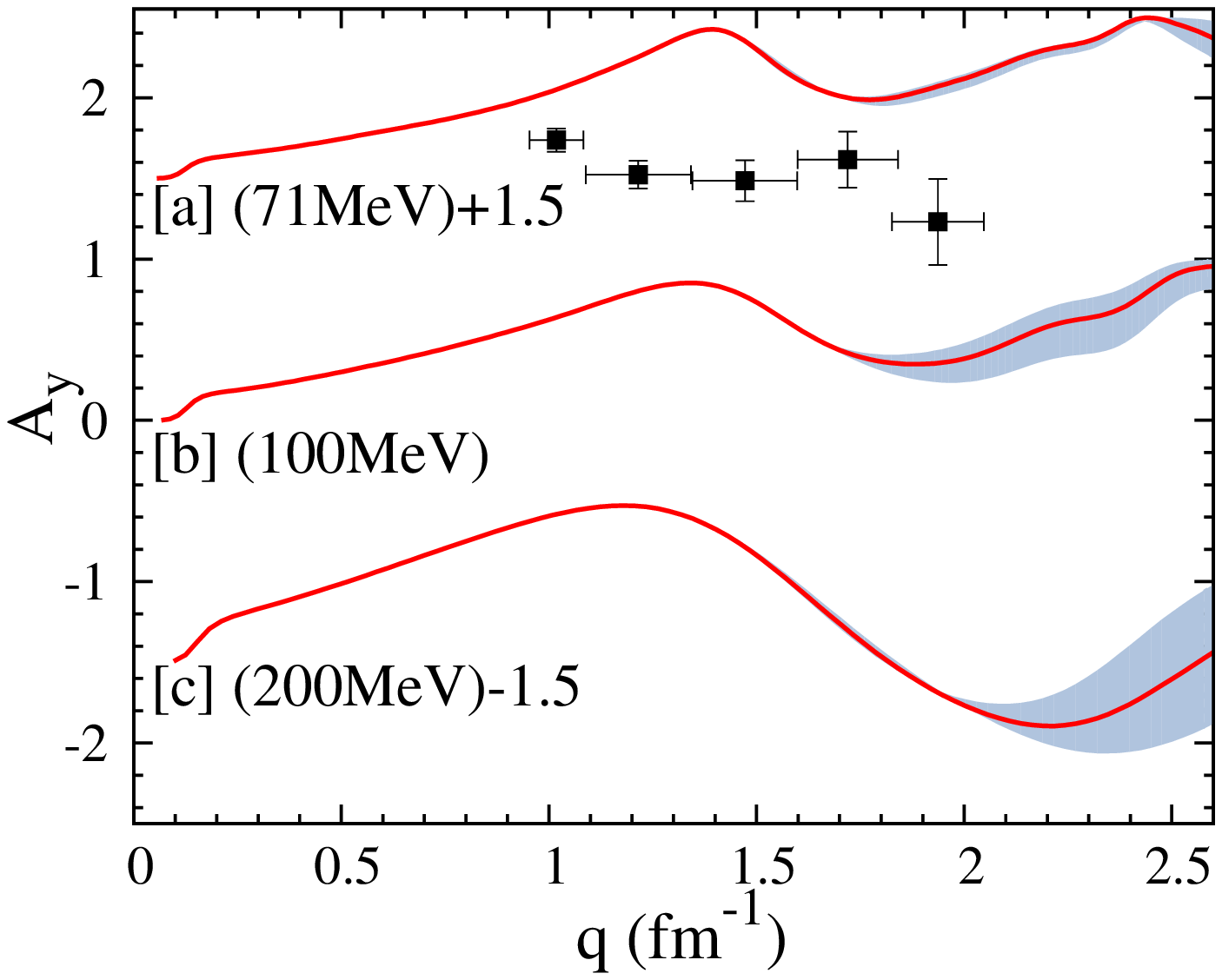}
\caption{(Color online) The angular distribution of the analyzing power ($A_y$)
for elastic scattering of $^{6}$He at
projectile energies [a] 71~MeV/nucleon, [b] 100~MeV/nucleon, and [c]
200~MeV/nucleon as function of the momentum transfer. The values for [a] and [c]
are shifted as indicated in the figure.  The meaning of
the lines is the same as in Fig.~\ref{fig9}. The data are taken from
Ref.~\cite{Uesaka:2010mm,Sakaguchi:2011rp}.
\label{fig10}
}
\end{center}
\end{figure}

The sensitivity to the variation in $\langle r^2_{ch}\rangle$, which translates to a variation
of $\nu_s$ is shown in Fig.~\ref{fig9} for the differential cross section as function of momentum
transfer for the different scattering energies. 
Since the difference between the measured values of the charge radius is quite small, the
variations in the differential cross section are also quite small. Since the charge radius also
enters the relation of the parameters $\nu_s$ and $\nu_p$ and the matter radius $\langle
r^2_{mat}\rangle$, Eq.~(\ref{eq:2.17}), we keep the matter radius constant at 2.33~fm. The
angular distribution of the analyzing power is shown in Fig.~\ref{fig10} as function of the
momentum transfer for the same three scattering energies. Here a larger sensitivity to the 
size of the charge radius is visibly for momentum transfers $q \ge 2 \rm{fm}^{-1}$ as indicated
by the shaded region. The sensitivity is larger for the higher scattering energies, indication
that at those energies more of the interior, i.e. the alpha-core, of $^6$He is probed.
Nevertheless, the variations are relatively small even at 200~MeV/nucleon, and probably not
experimentally accessible. 

\begin{figure}[htbp]
\begin{center}
\includegraphics[scale=.75]{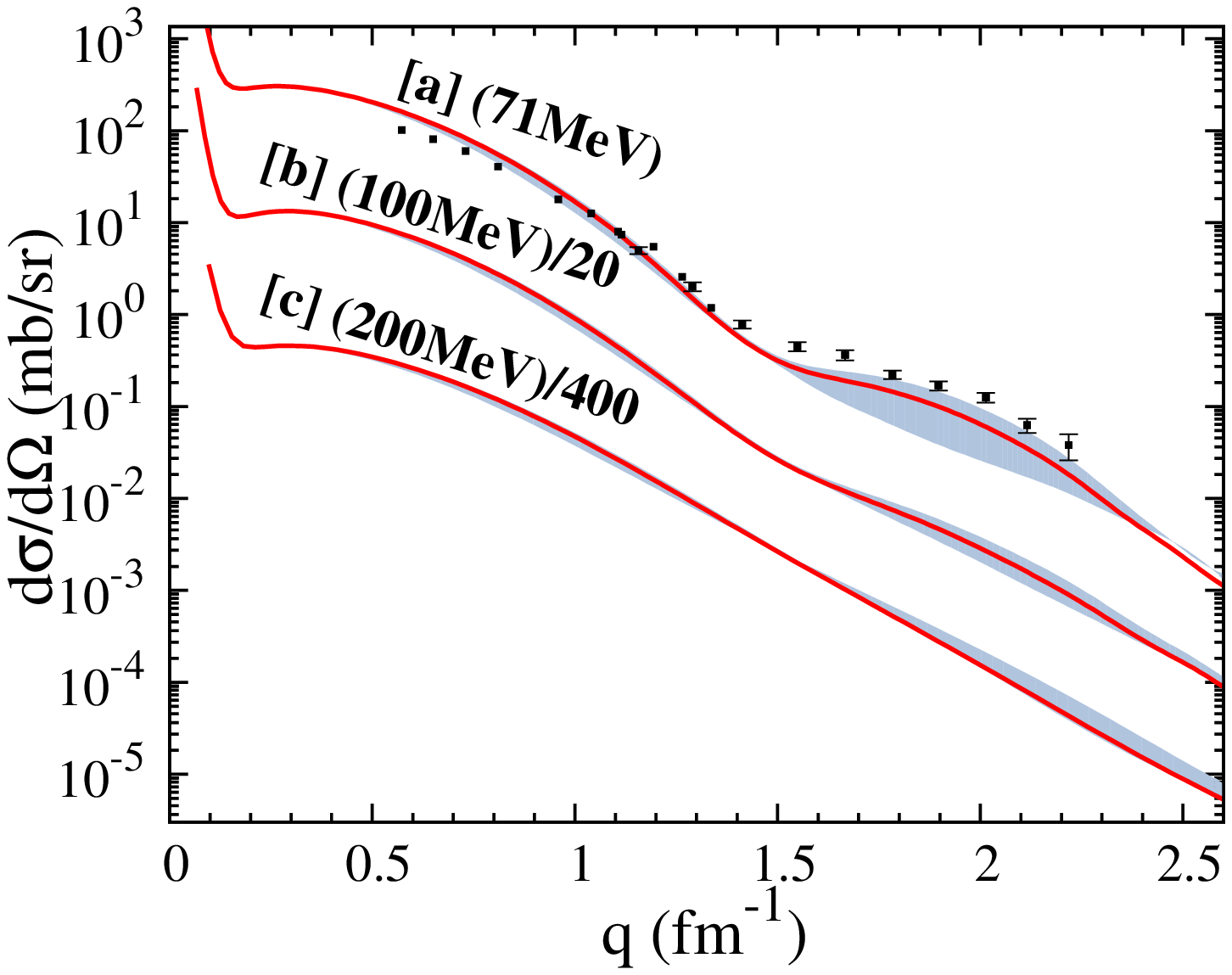}
\caption{(Color online) The angular distribution of the differential cross
section, $\frac{d\sigma}{d\Omega}$, for elastic scattering of $^{6}$He at
projectile energies [a] 71~MeV/nucleon, [b] 100~MeV/nucleon/40, and [c]
200~MeV/nucleon/400 as function of the momentum transfer. The solid (red) line
corresponds to the same calculation as the solid line in Fig.~\ref{fig3}, which 
which gives a charge radius $r_{ch}$~=~1.955~fm and a matter radius
$r_{mat}$~=~2.33~fm. The shaded region shows the variation of the matter radius
from 2.24~fm to 2.60~fm. 
The data are taken from
Refs.~\cite{Uesaka:2010mm,Korsheninnikov:1997mm}
\label{fig11}
}
\end{center}
\end{figure}

\begin{figure}[htbp]
\begin{center}
\includegraphics[scale=.75]{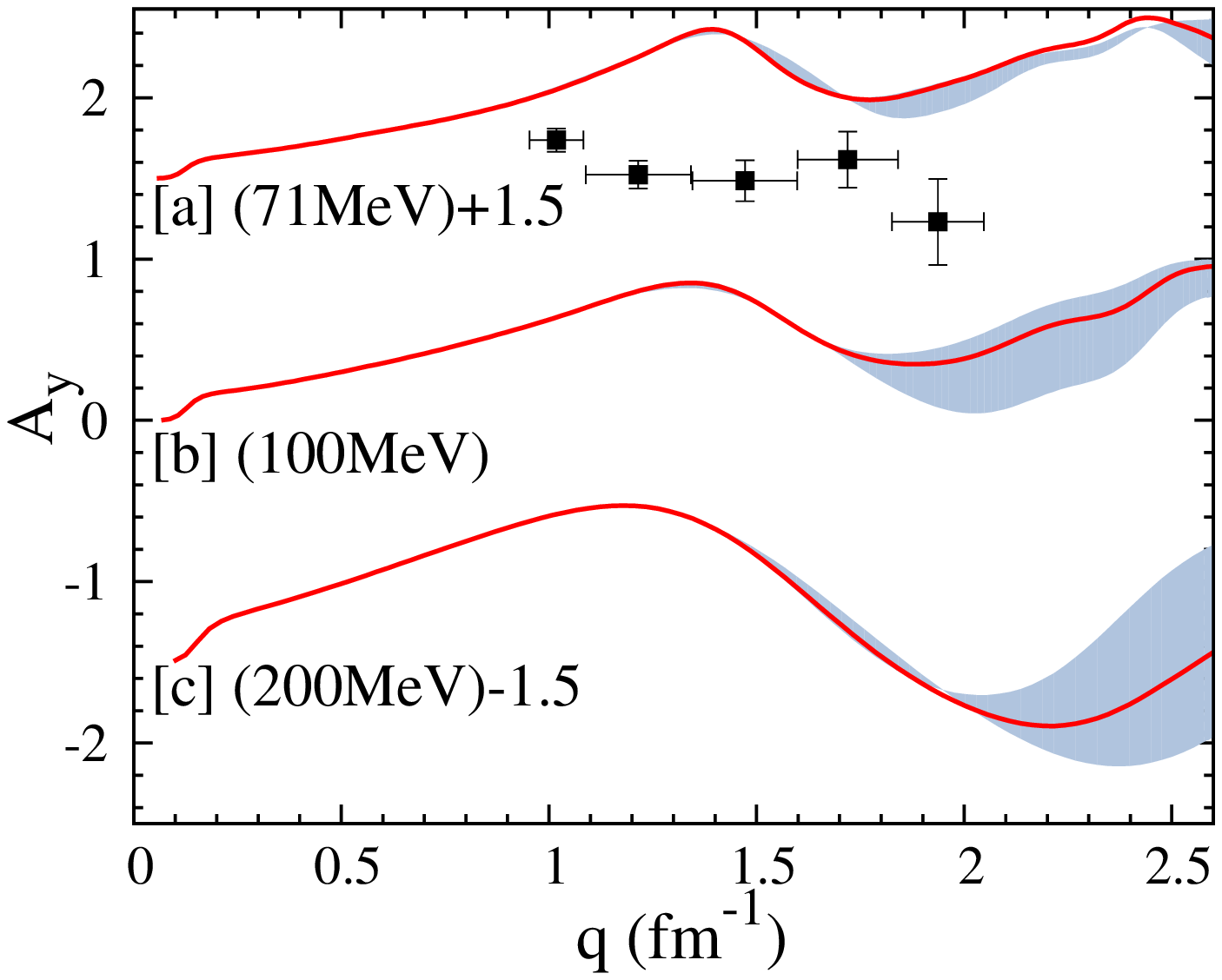}
\caption{(Color online) The angular distribution of the analyzing power ($A_y$)
for elastic scattering  of $^{6}$He at
projectile energies [a] 71~MeV/nucleon, [b] 100~MeV/nucleon, and [c]
200~MeV/nucleon as function of the momentum transfer. The values for [a] and [c]
are shifted as indicated in the figure.  The meaning of
the lines is the same as in Fig.~\ref{fig11}. The data are taken from
Ref.~\cite{Uesaka:2010mm,Sakaguchi:2011rp}.
\label{fig12}
}
\end{center}
\end{figure}

The matter radius is an extracted quantity and less well known than the charge radius. 
For testing the sensitivity of the scattering observables to the matter radius we keep the
charge radius fixed at 1.995~fm. As lower limit for the matter radius we choose the value of
2.24~fm extracted in Ref.~\cite{Bacca:2012up} and as upper limit the value of 2.6~fm used in
Ref.~\cite{Kaki:2012hr}. The sensitivity of the differential cross section  to the variation of
the matter radius in these limits is shown in Fig.~\ref{fig11} for three different scattering
energies. Here it is interesting to note that the lowest scattering energy, 71~MeV / nucleon,
shows the strongest sensitivity in the region between 1.5 and 2~fm$^{-1}$, indicated by the
shaded region. This most likely
results from the fact that the matter radius is dominated by the two outer valence neutrons.
The figure further indicates as far as our model is concerned, the data favor the smaller
values of the matter radius.  In Fig.~\ref{fig12} we show the sensitivity 
of the analyzing power to the same variation of the matter radius. It is interesting to
observe, that the analyzing power is less sensitive to the variation of the matter radius than
the differential cross section. However, this may be an artefact of our model, which puts the
to valence neutrons into the p$_{3/2}$-shell.  Again the two higher energies show considerably
more sensitivity to variations in the matter radius for momentum transfers $q \ge 2
\rm{fm}^{-1}$ as indicated by the shaded region in Fig.~\ref{fig11}.

\section{Open Shell Effects in the Optical Potential in $^8$He}

\begin{figure}[htbp]
\begin{center}
\includegraphics[scale=.75]{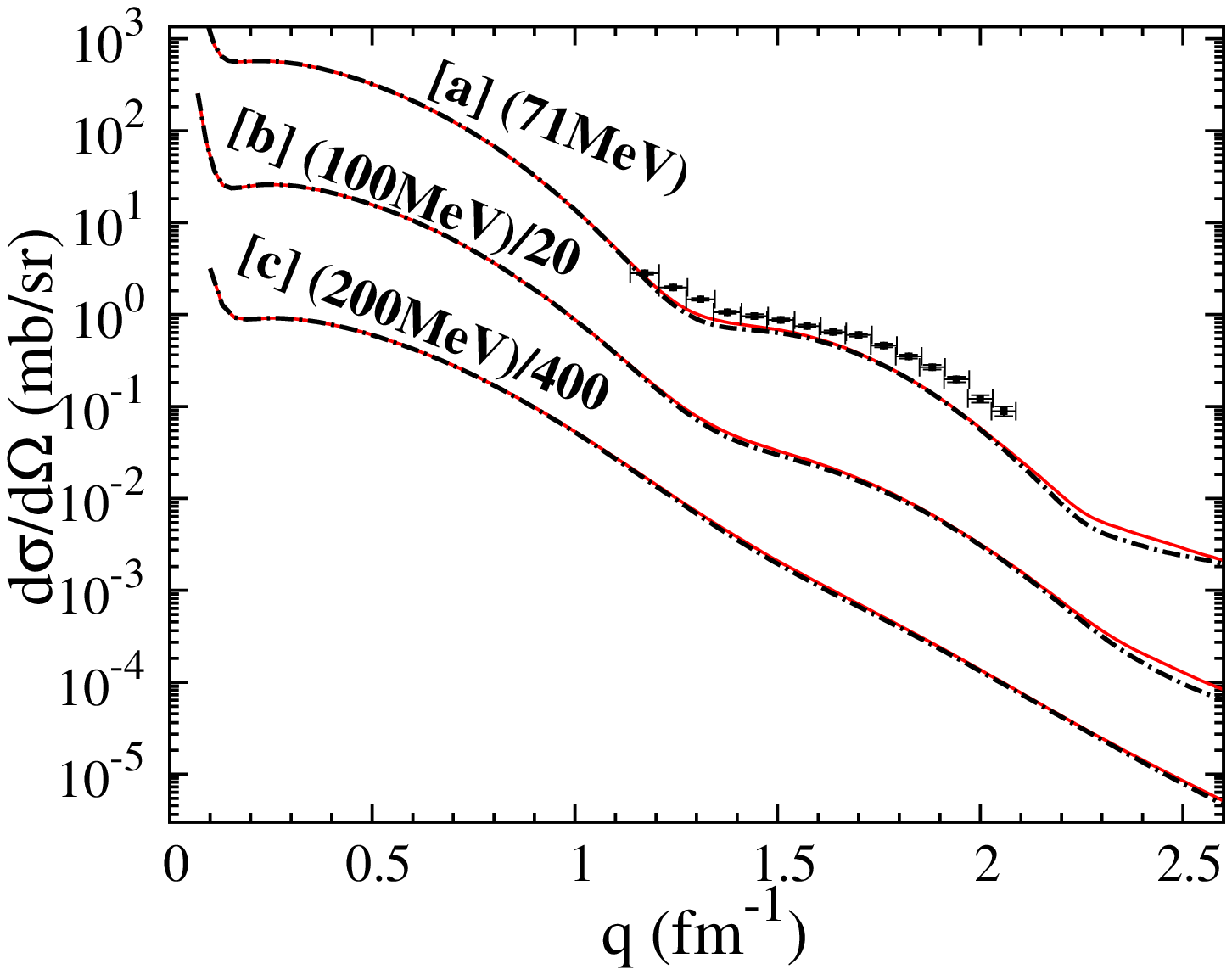}
\caption{(Color online) The angular distribution of the differential cross
section, $\frac{d\sigma}{d\Omega}$, for elastic scattering of $^{8}$He at
projectile energies [a] 71~MeV/nucleon, [b] 100~MeV/nucleon/40, and [c]
200~MeV/nucleon/400 as function of the momentum transfer. The calculations are
performed with an optical potential based on the CD-Bonn potential. The solid line
represents the full calculations, while the dash-dotted line represents the
calculating omitting open-shell effects. 
%The data are taken from Refs.~\cite{Uesaka:2010mm,Korsheninnikov:1997mm}
\label{fig13}
}
\end{center}
\end{figure}

\begin{figure}[htbp]
\begin{center}
\includegraphics[scale=.75]{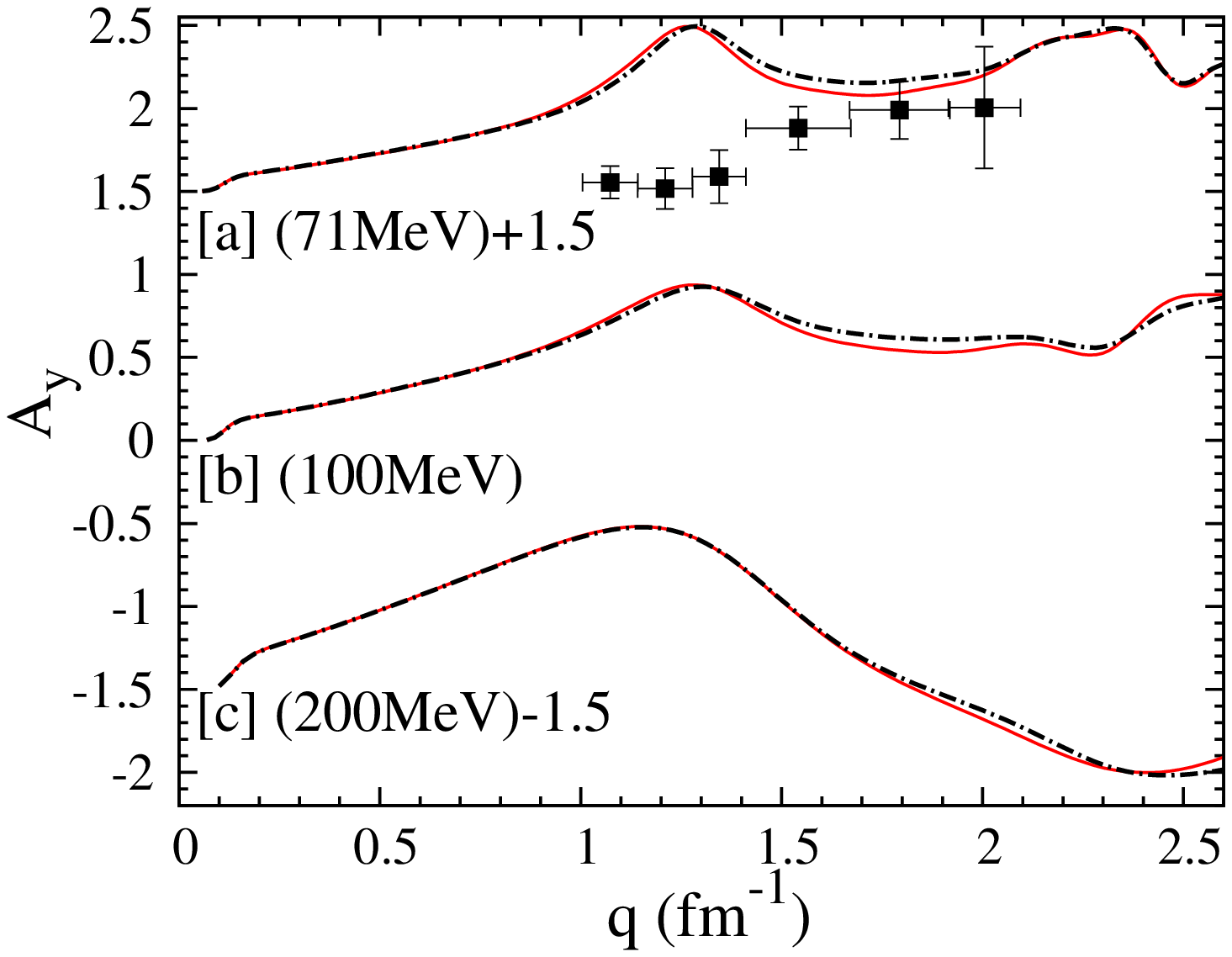}
\caption{(Color online) The angular distribution of the analyzing power ($A_y$)
for elastic scattering of of $^{8}$He at
projectile energies [a] 71~MeV/nucleon, [b] 100~MeV/nucleon, and [c]
200~MeV/nucleon as function of the momentum transfer. The values for [a] and [c]
are shifted as indicated in the figure.  The meaning of
the lines is the same as in Fig.~\ref{fig13}.
% The data are taken from Ref.~\cite{Uesaka:2010mm,Sakaguchi:2011rp}.
\label{fig14}
}
\end{center}
\end{figure}

The single particle density of $^6$He introduced in Section~\ref{subdenshe6} can be readily
extended to the single particle density of $^8$He. The p$_{3/2}$ shell can be occupied by four
valence neutrons coupled to total spin zero. Both helium isotopes have an alpha core, thus the
relation between the s-shell oscillator parameter $\nu_s$ and the charge radius of
Eq.~(\ref{eq:2.16}) is the same. The parameter $\nu_s$ determined from the
measured charge radius~\cite{Brodeur:2012zz} for $^8$He is given in Table~\ref{table-1}. The
relation between the matter radius and the parameters $\nu_s$ and $\nu_p$ is modified for $^8$He to
\begin{equation}
\langle r^2_{mat}\rangle = \frac{1}{8} \left( \frac{10}{\nu_p} + \frac{6}{\nu_s}\right).
\label{eq:4.1}
\end{equation}
Our calculations use the value of 2.53~fm from Ref.~\cite{Tanihata:1992wf} as matter radius.
Since in $^8$He the p$_{3/2}$-shell is occupied by double the amount of neutrons as the one in
$^6$He, one may speculate that the effect of the extra terms in the microscopic optical potential
resulting from these neutrons is larger compared to $^6$He. To investigate this we first
calculate the microscopic optical potential using only the terms generated by the Wolfenstein
amplitudes $A$ and $C$, and then compare to the corresponding calculations based on the
the expression of Eq.~(\ref{eq:2.3.1}). In Fig.~\ref{fig13} this comparison is shown for the
differential cross section for scattering of $^8$He off a proton target as function of the
momentum transfer for three selected energies. The effect of the additional terms in
Eq.~(\ref{eq:2.3.1}) are here vanishingly small. The corresponding comparison for the analyzing
power as function of the momentum transfer is depicted in Fig.~\ref{fig14}.  
The figure shows that the additional terms in the optical potential due the four valence neutrons
of $^8$He are in the same order of magnitude as shown in Fig.~\ref{fig4} for the analyzing power
of $^6$He. The reason may here be also that our model for the single-particle density with only
two oscillator wave functions is too simple.

\section{Summary and Conclusions}

In this work we extended the traditionally employed formulation of the first-order
microscopic optical
potential for elastic scattering from closed shell nuclei to nuclei with partially filled shells.
The complete full-folding integral for this first-order optical potential has been carried
out with the simplifying assumption that the single-particle density matrix for $^6$He and
$^8$He is given by a simple harmonic oscillator model. The alpha-core is described by a
single particle density matrix derived from one s-shell harmonic oscillator function, while the two valence
neutrons occupy the p$_{3/2}$-shell and are in the ground state coupled to spin zero. The
corresponding single particle density matrix is also derived from a single p-shell 
harmonic oscillator function. 

With these assumptions all terms of the optical potential that arise when 
integrating the six fully-off-shell Wolfenstein amplitudes of the NN scattering amplitude
with the single particle density matrix are derived and calculated. 
It turns out, that those Wolfenstein amplitudes that are related to the NN tensor force,
namely $G$, $H$, and $D$ do not contribute to optical potential when employing our model
ansatz for the single particle density matrix, in which the ground state consists of the two
valence neutrons occupying the p$_{3/2}$-shell. 
With our model single particle density the `traditional' first-order microscopic folding
optical potential, which consists of a central term related to the Wolfenstein amplitude
$A$ and a spin-orbit term related to the Wolfenstein amplitude $C$, acquires two new
additional terms. One of those terms is related to the Wolfenstein amplitude $C$, but since
it does not contain any spin-dependence, it adds to the central part of the optical
potential. The other term, which is related to the Wolfenstein amplitude $M$ adds to the
spin-orbit part of the optical potential. 

With these first-order folding optical potentials for $^6$He and $^8$He we calculated the
observables for elastic scattering, i.e. the differential cross section and the analyzing
power, at 71, 100, and 200~MeV per nucleon. We find that in all cases the additional terms
have a very small effect on the observables. 
This is most likely  result from the simplicity of our
model ansatz for the ground states of the two helium isotopes. Thus, we do not think it
appropriate to make a general conclusion about the importance of explicitly treating open
shell structure
in a microscopic optical potential. However, we would like to point out,
that our derivations open the path for employing sophisticated ground state wave
functions into a microscopic folding optical potential, as the ones provided by the
no-core shell-model~\cite{Navratil:2009ut,Cockrell:2012vd} (NSCM). 
In the NSCM the ground state of light nuclei is
calculated in a large $\hbar \Omega$ space. This leads to additional contributions
for each angular momentum state included in the NSCM. In addition, in a large $\hbar \Omega$
space transitions between different $l$ states will be allowed. Terms containing the
Wolfenstein amplitudes $G+H$ and $D$, which do not contribute in the simple $s$- and
$p_{3/2}$-shell model employed in this work, will contribute whenever $l\neq l'$ transitions
are included.
In this case all Wolfenstein amplitudes will contribute. 
As further remark, a NSCM single particle density matrix 
can be most naturally included in this formulation of the first-order microscopic folding
optical potential, since it is quite straightforward to derive a translationally invariant
single particle density using the NSCM~\cite{Navratil:2004dp}. 

We also want to point out that the formulation of a general
spin-dependent single particle density matrix of Section~\ref{subdenshe6} allows to consider
not only optical potential for the helium-isotopes as done in this work. The formulation
is written down for nuclear single particle densities with arbitrary spin.

Since $^6$He and $^8$He are both halo nuclei, with a small separation energy of the two
valence neutron, and thus a large spatial extension, we needed to investigate if our model
ansatz based on harmonic oscillator wave functions is inappropriate as input for the
optical potential. More specifically, we needed to investigate if an exponentially
decreasing spatial density, which is characteristic for halo nuclei, would yield
significantly different results for the scattering observables. We carried out this
investigation by matching an exponential tail at radii of about 3~fm to the oscillator
waver functions. The Fourier transform of these hybrid wave functions, 
after renormalization to the particle number, was used to derive single particle densities.
We find, that at the scattering energies under consideration,  the observables are
not sensitive to the long-range tail of the wave functions of the valence neutrons. 
This is a very encouraging result for plans to use no-core shell-model single particle
densities in calculating first-order optical potentials. 

Last, we performed a sensitivity study of the scattering observables to the charge and
matter radii of $^6$He. The charge radius of $^6$He is experimentally quite well known, and
thus when varying the s-wave oscillator parameter within the boundaries dictated by
experiment, we did not find a large variation in the observables. The situation is slightly
different for the matter radius, since this is often an extracted quantity, and we had
a larger range of variation. We found that the differential cross section at 71 MeV per
nucleon preferred a matter radius towards the smaller side of the values we considered.
The analyzing power at 100 and 200~MeV per nucleon shows sensitivity with respect to the
matter radius for momentum transfers $q \ge 2~{\rm fm}^{-1}$. The planned experiment at RIKEN
at this energy may be able to reach a momentum transfer of that size.

%-------------------------------------------------------------------------------
%****************************************************************************

%\newpage

\appendix

\section{Calculation of the Expectation Values}
\label{appendixA}

In this appendix we give some details  of the evaluation of the
spin-momentum operator of the scattering amplitude ${\overline M}$ of
Eq.~\ref{eq:2.5} in the target intrinsic frame. For the evaluation we
define the spin operator as
\begin{equation}
\bsigma \equiv (\sigma_1, \sigma_2, \sigma_3 ) \equiv
(\frac{\sigma_+-\sigma_-}{2i},\sigma_3,\frac{\sigma_++\sigma_-}{2}),
\label{eq:a.1}
\end{equation}
where the superscript $i$ is omitted since only the struck target nucleon is considered, and
$\sigma_\pm = \sigma_1\pm i\sigma_2$.
As indicated in Ref.~\cite{Elster:1989en}, in case of closed shell nuclei, the sum over
all states leads to a zero contribution of the spin-momentum operators. 
Considering the explicit expression of Eq.~(\ref{eq:2.9b}) for the
 p$_{3/2}$ wave function of the two valence neutrons coupled to total spin zero, 
we obtain when only considering the angular momentum parts
\begin{eqnarray}
\Phi_p ({\hat {\vvm p}}')\;\bsigma^{(i)}\cdot{\hat {\vvm n}}_{TI}\;\Phi_p
({\hat{\vvm p}})  \cr
&=& \frac{1}{12} \Big(
\frac{\hat{n}_3}{\sqrt{2}}\left(Y_1^0({\hat {\vvm p}}')\left\{Y_1^1({\hat {\vvm
p}})+Y_1^{-1}({\hat {\vvm p}})\right\}
-\left\{ Y_1^1({\hat {\vvm p}}')+Y_1^{-1}({\hat {\vvm p}}')\right\}Y_1^0({\hat {\vvm
p}})\right)\cr
&+& 2\hat{n}_2\left(Y_1^1 ({\hat {\vvm p}}')Y^{-1}_{1}({\hat {\vvm p}})
  -Y^{-1}_{1}({\hat {\vvm p}}') Y_1^1 ({\hat {\vvm p}}) \right) \cr
&+& \frac{i\hat{n}_1}{\sqrt{2}}\left(Y_1^0({\hat {\vvm p}}')\left\{Y_1^{-1}({\hat {\vvm
p}})-Y_1^1({\hat {\vvm p}}\right\}
+\left\{ Y_1^1({\hat {\vvm p}}')-Y_1^{-1}({\hat {\vvm p}}')\right\}Y_1^0({\hat
{\vvm p}}) \right) \Big) 
\label{eq:a.2}
\end{eqnarray}
The same form of expression is obtained when replacing ${\hat {\vvm n}}_{TI}$ 
with ${\hat {\vvm q}}_{TI}$ and ${\hat {\vvm P}}_{TI}$. Eq.~(\ref{eq:a.2}) is obtained
in the target intrinsic frame, which can be oriented arbitrarily with respect to other
frames. Therefore it is necessary to integrate over all possible orientations of the
target frame relative to the nucleon-nucleus frame, i.e. evaluate 
\begin{equation}
I = \frac{1}{8\pi^2} \int d{\hat {\vvm p}} {\hat {\vvm p}}' \Phi_p ({\hat {\vvm
p}}')\;\bsigma^{(i)}\cdot{\hat {\vvm n}}_{TI}\;\Phi_p
({\hat{\vvm p}}) \; \delta( {\hat {\vvm p}}\cdot {\hat {\vvm p}}' - \cos \alpha_{pp'}),
\label{eq:a.3}
\end{equation}
where the factor $8\pi^2$ is the norm of the integral with respect to a fixed angle
between the vectors ${\hat {\vvm p}}$ and ${\hat {\vvm p}}'$.
The delta function keeps the angle between ${\hat {\vvm p}}$ and ${\hat {\vvm p}}'$
fixed and can be expressed as
\begin{equation}
\cos\alpha_{pp'} = \cos\theta'\cos\theta+\sin\theta'\sin\theta\; \cos(\phi-\phi')
\label{eq:a.4}
\end{equation}
When the angle is fixed for a given ${\hat {\bf p}}'$, allowed orientations of the unit vector
${\hat {\vvm p}}$ form a cone. The projection of the cone's base onto the $xy$ plane is an ellipse
centered at
\begin{eqnarray}
X_c &= &\sin\theta'\cos\phi'\cos\alpha_{pp'} \cr
Y_c&=&\sin\theta'\sin\phi'\cos\alpha_{pp'}.
\label{eq:a.5}
\end{eqnarray}
With the major and minor axes given as $a=\sin \alpha_{pp'}$ and
$b=\cos\theta'\sin\alpha_{pp'}$ the parametric equation of the ellipse is determined
as 
\begin{eqnarray}
x&=&X_c + a\;\cos t\cos(\pi/2+\phi')-b\;\sin t\sin(\pi/2+\phi') \cr
y&=&Y_c + a\;\cos t\sin(\pi/2+\phi')+b\;\sin t\cos(\pi/2+\phi').
\label{eq:a.6}
\end{eqnarray}
The spherical harmonics depend on the angles $\theta$ and $\phi$, thus the integration
over the solid angle $\Omega$ can be replaced by the integration over the parameter $t$,
\begin{equation}
\int d\Omega \; \int d\Omega'\; Y^{m_1}_{m_1}({\hat {\vvm p}}') Y{m_2}_{m_1} ({\hat {\vvm
p}})\;\delta({\hat {p}} \cdot {\hat {p}}' -\cos\alpha_{pp'})
=\int d\Omega' \int\limits_{0}^{2\pi} dt Y^{m_1}_{m_1}({\hat {\vvm p}}') Y{m_2}_{m_1}
({\hat {\vvm p}}).  Y_1^{m2}({\hat {\vvm p}})
\label{eq:a.7}
\end{equation}
Substituting Eqs.~(\ref{eq:a.2}) and (\ref{eq:a.3}) and integrating leads to 
\begin{equation}
{\tilde \rho}_p(\vvm p, \vvm p')= -i\:\frac{2}{9}\frac{1}{\sqrt{\pi^3 \nu_p^5}}\;|
\vvm p \times\vvm p'| \; e^{-\frac{p^2+p'^2}{2\nu_p}},
\label{eq:a.8}
\end{equation}
which leads to Eq.~(\ref{eq:2.2.6}) after transforming to the variable ${\bf q}$ and {\bf
P}. 

For calculating the expectation value of $\bsigma^{(i)} \cdot {\hat {\vvm q}}$ the same
procedure is applied. Here we only have to consider that 
\begin{equation}
|q|=\sqrt{p^2+p'^2-2|p||p'|\cos\alpha_{pp'}}
\label{eq:a.9}
\end{equation}
and the unit vector ${\hat {\bf q}}$ as function of the angles $(\theta, \phi)$ and
$(\theta ', \phi ')$ is given as
\begin{equation}
{\hat {\vvm q}} = \frac{1}{|q|} \left(\begin{array}{cc}
 |p|\sin\theta\cos\phi-|p'|\sin\theta'\cos\phi'  \\
  |p|\sin\theta\sin\phi-|p'|\sin\theta'\sin\phi \\
  |p|\cos\theta-|p'|\cos\theta' \end{array}
 \right).
\label{eq:a.10}
\end{equation}
Inserting this into the corresponding integral, Eq.~(\ref{eq:a.3}) leads to
\begin{equation}
\frac{1}{8\pi^2} \int d{\hat {\vvm p}} d{\hat {\vvm p}}' 
\Phi_p ({\hat {\vvm p}}')\;\bsigma^{(i)}\cdot{\hat {\vvm q}}\;\Phi_p
({\hat{\vvm p}}) \; \delta( {\hat {\vvm p}}\cdot {\hat {\vvm p}}' - \cos \alpha_{pp'})
=0
\label{eq:a.11}
\end{equation}
The same integral for ${\bf P}$ also gives a zero contribution.

%%%%%%%%%%%%%%%%%%%%%%%%%%%%%%%%%%%%%%%%%%%%%%%%%%%%%%%%%%%%%%%%%%%%

\section{Explicit Calculation  of Contribution from the Wolfenstein amplitudes G+H, and D}
\label{appendixB}
As indicated in Eqs.~(\ref{eq:2.2.8}) and (\ref{eq:2.2.9}), the contributions of the Wolfenstein
amplitudes $G+H$ and $D$ vanish. In this appendix the explicit calculation is given. The structure
of the different terms in Eq.~(\ref{eq:2.2.8}) can be summarized as
\begin{eqnarray}
  U_1&=&\int d^3 P\;(G+H)\;\frac{1}{|K_{NN}|}\;|k_{NN}|\;\cos\alpha \;\tilde\rho(\vvm q,\vvm
P)\cr
 U_2&=&\int d^3 P\;(G+H)\;\frac{1}{|K_{NN}|}\;|k_{NN}'|\cos\gamma_{NN}\;\cos\alpha\;\tilde\rho(\vvm q,\vvm
P)\cr
 U_3&=&\int d^3 P\;(G+H)\;\frac{1}{|K_{NN}|}\;|k_{NN}'|\sin\gamma_{NN}\;\cos\alpha\;\tilde\rho(\vvm q,\vvm
P)
\label{eq:b1}
\end{eqnarray}
and for Eq.~(\ref{eq:2.2.9}) as
\begin{eqnarray}
 U_4&=&\int d^3 P\;D\;\frac{1}{|q|}\;|k_{NN}|\;\cos\alpha \;\tilde\rho(\vvm q,\vvm P)\cr
 U_5&=&\int d^3 P\;D\;\frac{1}{|q|}\;|k_{NN}'|\cos\gamma_{NN}\;\cos\alpha\;\tilde\rho(\vvm q,\vvm P)\cr
 U_6&=&\int d^3 P\;D\;\frac{1}{|q|}\;|k_{NN}'|\sin\gamma_{NN}\;\cos\alpha\;\tilde\rho(\vvm q,\vvm P),
\label{eq:b2}
\end{eqnarray}
where the integration explicity reads $\int d^3 P = \int_0^\infty P^2 dP \int_{-1}^1 d\cos
\theta_P
\int_0^{2\pi} d\phi_P$. We can show that the integrands are odd functions of the azimuthal angle
$\phi_P$, and thus the integrals in Eq.~(\ref{eq:b1}) and (\ref{eq:b2}) vanish.
In order to show this, we first note that $\tilde\rho(\vvm q,\vvm P)$ from Eq.~(\ref{eq:2.2.6}) contains
the cross product $|{\bf q}\times {\bf P}|$ and thus depends on $\sin \theta_P$.
Next we need to explicitly consider the term in  Eq.~(\ref{eq:b1}),
The magnitude of the vector $\vvm K_{NN}$ is given by
\begin{equation}
|K_{NN}|=\frac{1}{2}\sqrt{P^2+\left(\frac{A+1}{A}\right)^2 K^2-2\frac{A+1}{A}|P||K|\cos\gamma_{PK}}.
\label{eq:b3}
\end{equation}
Applying the addition theorem of spherical harmonics for $l=1$,  $\cos\gamma_{PK}$ can be
expressed as
\begin{equation}
\cos\gamma_{PK}=\cos\theta_P\cos\theta_{K}+\sin\theta_P\sin\theta_{K}\cos\phi_P.
\label{eq:b4}
\end{equation}
Thus, Eq.~(\ref{eq:b3}) can be re-expressed as
\begin{equation}
 |K_{NN}|=\sqrt{a-b\cos\phi_P},
\label{eq:b5}
\end{equation}
with
\begin{eqnarray}
  a&=&\frac{1}{4}\left(P^2+\left(\frac{A+1}{A}\right)^2
K^2-2\frac{A+1}{A}|P||K|\cos\theta_P\cos\theta_{K}\right)\cr
  b&=&\frac{1}{2}\frac{A+1}{A}|P||K|\sin\theta_P\sin\theta_{K}.
\label{eq:b6}
\end{eqnarray}
The angle $\cos\alpha$ which occurs in transformation between the NN and the target intrinsic frame
is defined as
\begin{equation}
 \cos\alpha  \equiv \widehat{\vvm q\times \vvm
P}\cdot\hat{\mathbf{K}}_{NN}=\frac{1}{2\sin\theta_P}\frac{A+1}{A}\frac{1}{|q||P||K_{NN}|}\sum_{i=1}^3\sum_{j=1}^3\sum_{k=1}^3\;\epsilon_{ijk}\;K_{i}\;q_{j}\;P_{k}.
\label{eq:b7}
\end{equation}
Choosing the reference frame such that ${\bf q}$ points along the z-axis, 
\begin{equation}
 \vvm q =(0,\;0,\;|q|);\;\;\;\vvm K=(K_{x},\;0,\;K_{z});\;\;\;\vvm P=(P_x,\;P_y,\;P_z),
\label{eq:b8}
\end{equation}
one obtains for Eq.~(\ref{eq:b7})
\begin{equation}
 \cos\alpha=-\frac{|K|\sin\theta_{K}}{2}\frac{A+1}{A}\frac{\sin\phi_P}{\sqrt{a-b\cos\phi_P}},
\label{eq:b9}
\end{equation}

The magnitudes of the vectors $\vvm k_{NN}$ and $\vvm k_{NN}'$ are given as
 \begin{eqnarray}
 | k_{NN}'|&=&\sqrt{K_{NN}^2+\frac{q^2}{4}+|K_{NN}||q|\cos\theta_{K_{NN}}}\cr
 | k_{NN}|&=&\sqrt{K_{NN}^2+\frac{q^2}{4}-|K_{NN}||q|\cos\theta_{K_{NN}}},
\label{eq:b10}
\end{eqnarray}
with
\begin{equation}
 \cos\theta_{K_{NN}}=\frac{\vvm q\cdot \vvm K_{NN}}{|q||K_{NN}|}=\frac{1}{2|K_{NN}||q|}\:\vvm q\cdot
\left(\frac{A+1}{A}\vvm K-\vvm P\right)
\equiv \frac{c}{\sqrt{a-b\cos\phi_P}},
\label{eq:b11}
\end{equation}
where
\begin{equation}
 c=\frac{1}{2}\left(\frac{A+1}{A}|K|\cos\theta_{K}-|P|\cos\theta_P\right).
\label{eq:b12}
\end{equation}
Introducing the abbreviations
\begin{eqnarray}
a_1'&=&a+\frac{q^2}{4}+\frac{|q|}{2}\left(\frac{A+1}{A}|K|\cos\theta_{K}-|P|\cos\theta_P\right)\cr
 a_1&=&a+\frac{q^2}{4}-\frac{|q|}{2}\left(\frac{A+1}{A}|K|\cos\theta_{K}-|P|\cos\theta_P\right),
\label{eq:b13}
\end{eqnarray}
Eq.~(\ref{eq:b10}) can be re-expressed as
\begin{eqnarray}
 | k_{NN}'|&=&\sqrt{a_1'-b\cos\phi_P}\cr
  | k_{NN}|&=&\sqrt{a_1-b\cos\phi_P}.
\label{eq:b14}
\end{eqnarray}
The angle $\cos \gamma_{NN}$ is given by
\begin{eqnarray}
 \cos\gamma_{NN}&\equiv &\frac{\vvm k_{NN}\cdot\vvm
k_{NN}'}{|k_{NN}||k_{NN}'|}=\frac{a_2-b\cos\phi_P}{\sqrt{a_1'-b\cos\phi_P}\;\sqrt{a_1-b\cos\phi_P}},
\label{eq:b15}
\end{eqnarray}
where $a_2=a-\frac{q^2}{4}$. Since $\sin \gamma_{NN}$ is obtained from $\cos \gamma_{NN}$,
both functions depend on $\cos\phi_P$ and thus are even with respect to  $\phi_P$.  

The functional dependence of the Wolfenstein amplitudes $G$, $H$, and $D$ are explicitly given by e.g.
\begin{equation}
G(\vvm q, \vvm K, \mathcal E) \equiv G(|q|,|K_{NN}|,\;\cos\theta_{K_{NN}},\;\mathcal E) =
G (\left(|q|,\;\sqrt{a-b\cos\phi_P},\;\frac{c}{\sqrt{a-b\cos\phi_P}},\mathcal E\right)\equiv G(\phi_P).
\label{eq:b16}
\end{equation}
Here we only give $G$. The functional dependence of $H$ and $D$ is exactly the same.
Considering the symmetry properties of $G$, Eq.~(\ref{eq:b16}), we see that $G(\pi + \phi_P) =
G(\pi- \phi_P)$. Thus when considering only the azimuthal part of the integration we obtain for
$U_1$ of Eq.~(\ref{eq:b1})
\begin{eqnarray}
  U_1&=&\int\limits_0^{2\pi} d\phi_P\;(G+H)\;\frac{\cos\alpha}{|K_{NN}|}\;|k_{NN}| \cr
 &=&\int\limits_0^{2\pi}
d\phi_P\;(G+H) \left(|q|,\;\sqrt{a-b\cos\phi_P},\;\frac{c}{\sqrt{a-b\cos\phi_P}},\mathcal
E\right)\;\frac{\sqrt{a_1-b\cos\phi_P}}{a-b\cos\phi_P}\sin\phi_P.
\label{eq:b17}
\end{eqnarray}

In this integration, for every point $\pi-\phi_P$ there is another point $\pi+\phi_P,$ with the same
value of $\cos\phi_P$. This means,  the  Wolfenstein amplitudes $G$, $H$, and $D$  have identical
values at the points $\pi\pm\phi_P$. On the other hand, the $sine$ function is odd with respect to $\phi_P$.
Therefore, the contribution of each point $\pi-\phi_P$ to the integral is canceled by the contribution of
the point at $\pi+\phi_P$. Consequently, the overall integral is zero. 
The same argument applies to all other functions $U_i$ of Eqs.~(\ref{eq:b1}) and (\ref{eq:b2}), which 
leads to the result that all integrals give zero, thus concluding our proof.

%%%%%%%%%%%%%%%%%%%%%%%%%%%%%%%%%%%%%%%%%%%%%%%%%%%%%%%%%%%%%%%%%%%%%%%%%%%%%%%%%%%%

\begin{acknowledgments}
This work was performed in part under the
auspices of the U.~S.  Department of Energy under contract
No. DE-FG02-93ER40756 with Ohio University and
under contract No. DE-SC0004084 (TORUS Collaboration).  S.P.W. thanks the
Institute of Nuclear and Particle Physics (INPP) and the Department of
Physics and Astronomy at Ohio University for their hospitality and
support during his sabbatical stay. C.E. appreciates  clarifying and helpful discussions
about theoretical aspects of the work with R.C. Johnson.
\end{acknowledgments}

%%%%%%%%%%%%%%%%%%%%%%%%%%%%%%%%%%%%%%%%%%%%%%%%%%%%%%%%%%%%%%%%%%%%

\bibliography{clusters}

\clearpage
%%%%%%%%%%%%%%%%%%%%%%%%%%%%%%%%%%%%%%%%%%%%%%%%%%%%%%

\begin{table}
\begin{tabular}{|c|c|c|cc|} \hline \hline
     & $r_{ch}$ [fm] &  $r_{mat}$ [fm] & $\nu_s$ [fm$^2$] & $\nu_p$ [fm$^2$] \\
\hline \hline
$^6$He & 1.955~\cite{Mueller:2008bj} & 2.333~\cite{wang-thesis} & 0.393 & 0.289 \\
$^8$He & 1.885~\cite{Brodeur:2012zz} & 2.53~\cite{Tanihata:1992wf} & 0.422 & 0.270\\

\hline\hline
\end{tabular}
\caption{ The charge and matter radii $r_{ch}$ and $r_{mat}$ used to determine the
oscillator parameters for the density matrices of $^6$He and $^8$He.
}
\label{table-1} 
\end{table}

\begin{table}
\begin{tabular} {|c|c|c|c|c|c|c|}
\hline \hline
$R_{m}$ [fm] & s-shell [\%] & p-shell [\%] & $\mu$ [fm$^{-1}$] & B [fm$^{-3}$]
& norm (p-shell) & $r_{mat}$ [fm] \\
\hline\hline
2.8 & 89.6  &52.5   & 0.453 & 0.833 & 3.05 & 2.89 \\
3.2 & 95.5  &68.6   & 0.613 & 1.347 &2.35  & 2.55\\
3.5 & 97.8  &78.6   & 0.727 & 1.970& 2.17 & 2.44\\
3.8 & 99.0  &86.2   & 0.836 & 2.936 &2.08 & 2.39 \\
\hline \hline
\end{tabular}
\caption{Parameters for matching an exponential tail to the p-shell harmonic
oscillator wave function. The detailed explanation of the parameters is given
in Section~\ref{exptail}.}
\label{table-2}
\end{table}

\vfill

%%%%%%%%%%%%%%%%%%%%%%%%%%%%%%%%%%%%%%%%%%%%%%%%%%%%%%

\newpage

\noindent

%\vspace{10mm}

\vspace{10mm}

\vspace{20mm}

\end{document}